\documentclass[twocolumn]{IEEEtran}
\usepackage{amsmath}
\usepackage{amssymb}
\usepackage{amsfonts}
\usepackage{slashbox}
\usepackage{graphicx}
\usepackage{cite, amsmath, amsfonts, amssymb, psfrag, epsfig,tikz}
\usetikzlibrary{shapes,shadows,arrows}
\newtheorem{proposition}{{Proposition}}
\newtheorem{definition}{{Definition}}
\newtheorem{theorem}{{Theorem}}

\DeclareMathAlphabet{\mathpzc}{OT1}{pzc}{m}{it}

\begin{document}

\tikzstyle{line} = [draw, -stealth,semithick]
\tikzstyle{block} = [draw, rectangle, text width=7em, text centered, minimum height=12mm, node distance=8em,semithick]
\tikzstyle{chblock} = [draw, rectangle, text width=4em, text centered, minimum height=40mm, node distance=8em,semithick]

\title{A Source-Channel Separation Theorem with Application to the Source Broadcast Problem}


\author{Kia Khezeli and Jun Chen\thanks{This work was supported in part by an
Early Researcher Award from the Province of Ontario and in part by the
Natural Science and Engineering Research Council (NSERC) of Canada
under a Discovery Grant. This paper was presented in part at the 2014 IEEE
International Symposium on Information Theory.
}
\thanks{K. Khezeli was with the Department
of Electrical and Computer Engineering, McMaster University,
Hamilton, ON L8S 4K1, Canada. He is now with the School of Electrical and Computer Engineering, Cornell University, Ithaca, NY 14853, USA (email: kk839@cornell.edu).}
\thanks{J. Chen is with the Department
of Electrical and Computer Engineering, McMaster University,
Hamilton, ON L8S 4K1, Canada  (email: junchen@ece.mcmaster.ca).}}

\maketitle

\begin{abstract}
A converse method is developed for the source broadcast problem. Specifically, it is shown that the separation architecture is optimal for a variant of the source broadcast problem and the associated source-channel separation theorem can be leveraged,  via a reduction argument, to establish a necessary condition for the original problem, which unifies several existing results in the literature.  Somewhat surprisingly, this method, albeit based on the source-channel separation theorem, can be used to prove the optimality of non-separation based schemes and determine the performance limits in certain scenarios where the separation architecture is suboptimal.


\end{abstract}


\section{Introduction}\label{sec:introduction}

In the source broadcast problem, a source is sent over a broadcast channel through suitable encoding and decoding so that the reconstructions at the receivers satisfy the prescribed constraints. The special case of sending a Gaussian source over a Gaussian broadcast channel has received particular attention. For this special case, it is known that source-channel separation  is in general suboptimal \cite{Goblick65} and hybrid digital-analog coding schemes can outperform pure digital/analog schemes \cite{MP02,RFZ06,NCW07,PPR11}. The extension of the hybrid coding architecture to the non-Gaussian setting can be found in \cite{MLK15}.

In contrast, the progress on the converse side is still somewhat limited. To the best of our knowledge, the first non-trivial result in this direction was obtained by Reznic \textit{et al.} \cite{RFZ06} for the scalar version of the aforementioned Gaussian case. The converse argument in \cite{RFZ06} involves an auxiliary random variable, which is generated by the source via an additive Gaussian noise channel. This auxiliary random variable is constructed in exactly the same manner as the one in Ozarow's celebrated work on the Gaussian multiple description problem \cite{Ozarow80}. However, this resemblance is, in a certain sense, rather superficial. Indeed, on a more technical level,  the auxiliary random variable introduced by Ozarow (as elucidated in \cite{WV07,WV09,Chen09,SSC13}) plays the role of  exploiting an implicit conditional independence structure whereas the role of the auxiliary random variable in \cite{RFZ06} is apparently different and still largely elusive.  Recent years have seen several new converse results \cite{TDS11a,TKS13,CTDS14} for the source broadcast problem. These results are based on arguments similar to the original one by Reznic \textit{et al.}, especially in terms of the way the auxiliary random variables are constructed and exploited. It is worth noting that such arguments can only handle a restricted class of auxiliary random variables (essentially those that can be generated by the source via certain additive noise channels); this restriction typically leads to certain constraints on the set of sources, channels, or distortion measures that can be analyzed.  

The present paper is, to a certain extent, an outcome of our effort in seeking a conceptual understanding of the converse argument by Reznic \textit{et al.} in general and the role of the associated auxiliary random variable in particular. We shall show that one can establish a source-channel separation theorem for a variant of the source broadcast problem and leverage it to derive a necessary condition for the original problem. This necessary condition, when specialized to the case of sending a scalar Gaussian source over a Gaussian broadcast channel, recovers the corresponding result by Reznic \textit{et al.} \cite{RFZ06}; moreover, in this way, the converse argument in \cite{RFZ06} finds a simple interpretation, and the associated auxiliary random variable acquires an operational meaning. It should be pointed out that, in our approach, the auxiliary random variable can be generated by the source in an arbitrary manner. Therefore, the restriction imposed in the existing arguments \cite{TDS11a,TKS13,CTDS14} is in fact unnecessary. On the other hand, the problem of identifying the optimal auxiliary random variable naturally arises due to this additional freedom. It will be seen that the analytical solutions for this problem can be found in some special cases; interestingly, these solutions indicate that the specific choices of auxiliary random variables in \cite{RFZ06,TKS13} are actually optimal in their respective contexts.

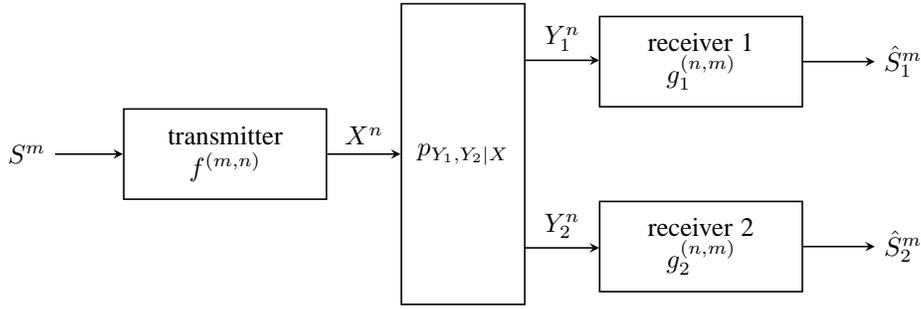
\begin{figure*}
\centering
\begin{tikzpicture}
\node [] (Source) {$S^m$};
\node [block, right of=Source , xshift=-0.5em] (Enc) {transmitter\\ $f^{(m,n)}$};
\node [chblock, right of=Enc , xshift=1em] (Channel) {$p_{Y_1,Y_2|X}$};\
\node [block, right of=Channel , xshift=1em, yshift=3.5em] (Dec1) {receiver 1\\ $g_1^{(n,m)}$};
\node [block, right of=Channel , xshift=1em, yshift=-3.5em] (Dec2) {receiver 2\\ $g_2^{(n,m)}$};
\node [right of=Dec1 , xshift=4.8em] (Rec1) {$\hat{S}_{1}^m$};
\node [right of=Dec2 , xshift=4.8em] (Rec2) {$\hat{S}_{2}^m$};
\path [line] (18.85em,3.55em)--node[xshift=0em,above] {$Y_{1}^n$}(21.65em,3.55em);
\path [line] (18.85em,-3.55em)--node[xshift=0em,above] {$Y_{2}^n$}(21.65em,-3.55em);
\path [line] (Source)--(Enc);
\path [line] (Enc) -- node[xshift=0em,above] {$X^n$} (Channel);
\path [line] (Dec1)--(Rec1);
\path [line] (Dec2)--(Rec2);
\end{tikzpicture}
\caption{System $\Pi$}\label{fig:systemA}
\end{figure*}

Our work is also partly motivated by the problem of sending a bivariate Gaussian source over a Gaussian broadcast channel first studied by Bross \textit{et al.} \cite{BLT10}. For this problem, it is known that the achievable distortion region of a certain hybrid digital-analog coding scheme \cite{TDS11} matches the outer bound in \cite{BLT10} whereas separate source-channel coding is in general suboptimal \cite{TDS11,GT13}.  An alternative proof of the outer bound in \cite{BLT10} was recently obtained by Song \textit{et al.} \cite{SCT15}. This new proof \cite{SCT15} bears some similarity to the aforementioned converse argument by Reznic \textit{et al.} \cite{RFZ06}. We will clarify their connection by giving a unified proof for the vector Gaussian case, which implies, among other things, that  the outer bound in \cite{BLT10} can be deduced from the general necessary condition for the source broadcast problem found in the present paper. Therefore, our converse method,  albeit based on the source-channel separation theorem,  can be used to prove the optimality of non-separation based schemes and determine the performance limits in certain scenarios where the separation architecture is suboptimal.



The rest of this paper is organized as follows. We present the problem setup  in Section \ref{sec:setup} and the relevant capacity results for broadcast channels with receiver side information in Section \ref{sec:broadcast}. We establish a source-channel separation theorem for a variant of the source broadcast problem in Section \ref{sec:separation}. It is shown in Section \ref{sec:application} that this separation theorem can be used in conjunction with a simple reduction argument to derive a necessary condition for the original source broadcast problem; moreover, this necessary condition is evaluated for the special case of the binary uniform source with the Hamming distortion measure. The quadratic Gaussian case is treated in Section \ref{sec:Gaussian}.  We conclude the paper in Section \ref{sec:conclusion}.

Throughout this paper, the binary entropy function and its inverse are denoted by $H_b(\cdot)$ and $H^{-1}_b(\cdot)$, respectively. For any $a,b\in[0,1]$, we define $a*b=a(1-b)+(1-a)b$.  The logarithm function is assumed to be base 2 unless specified otherwise.

\section{Problem Setup}\label{sec:setup}

 The source broadcast system (System $\Pi$) consists of the following components (see Fig. \ref{fig:systemA}):
\begin{itemize}
\item an i.i.d source $\{S(t)\}_{t=1}^\infty$ with marginal distribution $p_S$ over alphabet $\mathcal{S}$,

\item a discrete memoryless broadcast channel $p_{Y_1,Y_2|X}$ with input alphabet $\mathcal{X}$ and output alphabets $\mathcal{Y}_i$, $i=1,2$,

\item a transmitter, which is equipped with an encoding function $f^{(m,n)}:\mathcal{S}^m\rightarrow\mathcal{X}^n$ that maps a block of source samples $S^m\triangleq(S(1),\cdots,S(m))$ of length $m$ to a channel input block $X^n\triangleq(X(1),\cdots,X(n))$ of length $n$ (the number of channel uses per source sample, i.e., $\frac{n}{m}$, is referred to as the bandwidth expansion ratio),

\item two receivers, where receiver $i$ is equipped with 
    a decoding function $g^{(n,m)}_i:\mathcal{Y}^n_i\rightarrow\hat{\mathcal{S}}^m_i$ that  maps the channel output block $Y^n_{i}\triangleq(Y_i(1),\cdots,Y_i(n))$ generated by $X^n$  to a source reconstruction block $\hat{S}^m_{i}\triangleq(\hat{S}_i(1),\cdots,\hat{S}_i(m))$, $i=1,2$.
\end{itemize}
  Unless stated otherwise, we assume that $\mathcal{S}$, $\hat{\mathcal{S}}_1$, $\hat{\mathcal{S}}_2$, $\mathcal{X}$, $\mathcal{Y}_1$, and $\mathcal{Y}_2$ are finite sets.




Let $\mathcal{P}_{\mathcal{S}\times\hat{\mathcal{S}}_i}(p_S)$ denote the set of joint distributions over $\mathcal{S}\times\hat{\mathcal{S}}_i$ with the marginal distribution on $\mathcal{S}$ fixed to be $p_S$, $i=1,2$.


\begin{definition}\label{def:systemA}
 Let $\kappa$ be a non-negative number and $\mathcal{Q}_i$ be a non-empty compact subset of $\mathcal{P}_{\mathcal{S}\times\hat{\mathcal{S}}_i}(p_S)$, $i=1,2$. We say $(\kappa,\mathcal{Q}_1,\mathcal{Q}_2)$ is achievable for System $\Pi$ if, for every $\epsilon>0$, there exist encoding function $f^{(m,n)}:\mathcal{S}^m\rightarrow\mathcal{X}^n$ and decoding functions $g^{(n,m)}_i:\mathcal{Y}^n_i\rightarrow\hat{\mathcal{S}}^m_i$, $i=1,2$, such that
\begin{align}
&\frac{n}{m}\leq\kappa+\epsilon,\label{eq:maincond1}\\
&\min\limits_{q_i\in\mathcal{Q}_i}\left\|\frac{1}{m}\sum\limits_{t=1}^mp_{S(t),\hat{S}_i(t)}-q_i\right\|\leq\epsilon,\quad i=1,2,\label{eq:maincond2}
\end{align}
where $\|\cdot\|$ is the 1-norm.
The set of all achievable $(\kappa, \mathcal{Q}_1,\mathcal{Q}_2)$ for System $\Pi$ is denoted by $\Gamma$.
\end{definition}

Remark: It is easy to verify that
\begin{align*}
\frac{1}{m}\sum_{t=1}^mp_{S(t),\hat{S}_i(t)}\in\mathcal{P}_{\mathcal{S}\times\hat{\mathcal{S}}_i}(p_S),\quad i=1,2.
\end{align*}

Now consider the following more conventional definition.

\begin{definition}\label{def:systemA2}
 Let $w_i:\mathcal{S}\times\hat{\mathcal{S}}_i\rightarrow[0,\infty)$ be two distortion measures. For non-negative numbers $\kappa$, $d_1$, and $d_2$,  we say $(\kappa,d_1,d_2)$ is achievable for System $\Pi$ under distortion measures $w_1$ and $w_2$ if, for every $\epsilon>0$, there exist encoding function $f^{(m,n)}:\mathcal{S}^m\rightarrow\mathcal{X}^n$ and decoding functions $g^{(n,m)}_i:\mathcal{Y}^n_i\rightarrow\hat{\mathcal{S}}^m_i$, $i=1,2$, such that
\begin{align}
&\frac{n}{m}\leq\kappa+\epsilon,\nonumber\\
&\frac{1}{m}\sum\limits_{t=1}^m\mathbb{E}[w_i(S(t),\hat{S}_i(t))]\leq d_i+\epsilon,\quad i=1,2.\label{eq:maincond22}
\end{align}
\end{definition}

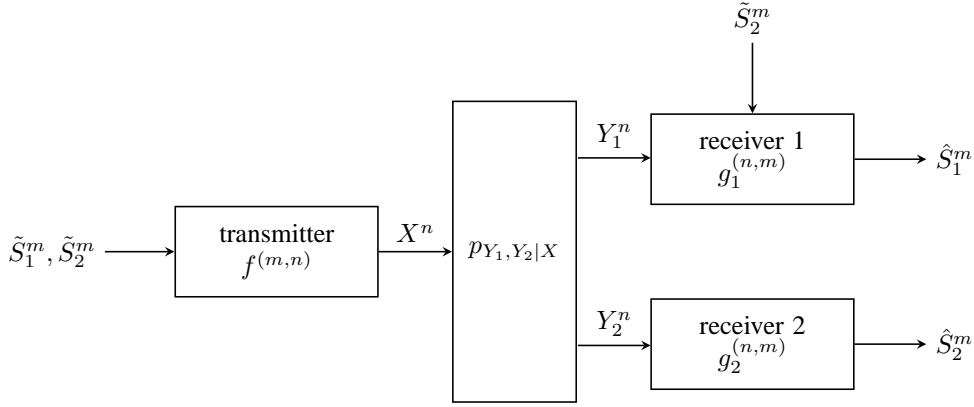
\begin{figure*}
\centering
\begin{tikzpicture}
\node [] (Source) {$\tilde{S}_{1}^m, \tilde{S}_{2}^m$};
\node [block, right of=Source , xshift=0.5em] (Enc) {transmitter\\ $f^{(m,n)}$};
\node [chblock, right of=Enc , xshift=1em] (Channel) {$p_{Y_1,Y_2|X}$};\
\node [block, right of=Channel , xshift=1em, yshift=3.5em] (Dec1) {receiver 1\\ $g_1^{(n,m)}$};
\node [above of=Dec1, yshift=2.5em] (Side) {$\tilde{S}_{2}^m$};
\node [block, right of=Channel , xshift=1em, yshift=-3.5em] (Dec2) {receiver 2\\ $g_2^{(n,m)}$};
\node [right of=Dec1 , xshift=4.8em] (Rec1) {$\hat{S}_{1}^m$};
\node [right of=Dec2 , xshift=4.8em] (Rec2) {$\hat{S}_{2}^m$};

\path [line] (19.9em,3.55em)--node[xshift=0em,above] {$Y_{1}^n$}(22.65em,3.55em);
\path [line] (19.9em,-3.55em)--node[xshift=0em,above] {$Y_{2}^n$}(22.65em,-3.55em);
\path [line] (Source)--(Enc);
\path [line] (Enc) -- node[xshift=0em,above] {$X^n$} (Channel);
\path [line] (Dec1)--(Rec1);
\path [line] (Dec2)--(Rec2);
\path [line] (Side)--(Dec1);
\end{tikzpicture}
\caption{System $\tilde{\Pi}$}\label{fig:systemB}
\end{figure*}

The following result shows that Definition \ref{def:systemA} is more general than Definition \ref{def:systemA2}.
\begin{proposition}\label{prop:equivalence}
$(\kappa,d_1,d_2)$ is achievable for System $\Pi$ under distortion measures $w_1$ and $w_2$ if and only if $(\kappa,\mathcal{Q}(w_1,d_1),\mathcal{Q}(w_2,d_2))\in\Gamma$, where $\mathcal{Q}(w_i,d_i)=\{p_{S,\hat{S}_i}\in\mathcal{P}_{\mathcal{S}\times\hat{\mathcal{S}}_i}(p_S):\mathbb{E}[w_i(S,\hat{S}_i)]\leq d_i\}$, $i=1,2$.
\end{proposition}
\begin{IEEEproof}
 Let $T$ be a random variable independent of $(S^m,\hat{S}^m_{1},\hat{S}^m_{2})$ and uniformly distributed over $\{1,\cdots,m\}$. It is easy to verify that (\ref{eq:maincond2}) can be written equivalently as
 \begin{align*}
 \min\limits_{q_i\in\mathcal{Q}_i}\left\|p_{S(T),\hat{S}_i(T)}-q_i\right\|\leq\epsilon,\quad i=1,2,
 \end{align*}
 and (\ref{eq:maincond22}) can be written equivalently as
 \begin{align*}
 \mathbb{E}[w_i(S(T),\hat{S}_i(T))]\leq d_i+\epsilon,\quad i=1,2.
 \end{align*}
Note that
 \begin{align*}
 &\mathbb{E}[w_i(S(T),\hat{S}_i(T))]\\
 &=\sum\limits_{s\in\mathcal{S},\hat{s}_i\in\hat{\mathcal{S}}_i}p_{S(T),\hat{S}_i(T)}(s,\hat{s}_i)w_i(s,\hat{s}_i)\\
 &\leq\sum\limits_{s\in\mathcal{S},\hat{s}_i\in\hat{\mathcal{S}}_i}q_i(s,\hat{s}_i)w_i(s,\hat{s}_i)\\
 &\quad+\sum\limits_{s\in\mathcal{S},\hat{s}_i\in\hat{\mathcal{S}}_i}|p_{S(T),\hat{S}_i(T)}(s,\hat{s}_i)-q_i(s,\hat{s}_i)|w_i(s,\hat{s}_i)\\
 &\leq d_i+\|p_{S(T),\hat{S}_i(T)}-q_i\|\max\limits_{s\in\mathcal{S},\hat{s}_i\in\hat{\mathcal{S}}_i}w_i(s,\hat{s}_i)
 \end{align*}
 for any $q_i\in\mathcal{Q}_i(w_i,d_i)$, $i=1,2$. Therefore, we have
 \begin{align*}
  &\mathbb{E}[w_i(S(T),\hat{S}_i(T))]\\
  &\leq d_i+\min\limits_{q_i\in\mathcal{Q}_i(w_i,d_i)}\|p_{S(T),\hat{S}_i(T)}-q_i\|\max\limits_{s\in\mathcal{S},\hat{s}_i\in\hat{\mathcal{S}}_i}w_i(s,\hat{s}_i),\\
  &\hspace{2.5in} i=1,2,
 \end{align*}
 from which the ``if" part follows immediately.

 Now we proceed to prove the ``only if" part. Assume that $(\kappa,d_1,d_2)$ is achievable for System $\Pi$ under distortion measures $w_1$ and $w_2$. For every $\epsilon>0$, according to Definition \ref{def:systemA2}, we can find encoding function $f^{(m,n)}:\mathcal{S}^m\rightarrow\mathcal{X}^n$ and decoding functions $g^{(n,m)}_i:\mathcal{Y}^n_i\rightarrow\hat{\mathcal{S}}^m_i$, $i=1,2$, satisfying $\frac{n}{m}\leq\kappa+\epsilon$ and $\mathbb{E}[w_i(S(T),\hat{S}_i(T))]\leq d_i+\epsilon$, $i=1,2$. We
 shall denote $S(T)$ simply by $S$ since the distribution of  $S(T)$ is $p_S$, and denote $\hat{S}_1$ and $\hat{S}_2$ by $\hat{S}^{(\epsilon)}_1$ and $\hat{S}^{(\epsilon)}_2$, respectively, to stress their dependence on $\epsilon$. Note that $\{p_{S,\hat{S}^{(\epsilon)}_1,\hat{S}^{(\epsilon)}_2}:\epsilon>0\}$ is contained in a compact set and $\mathbb{E}[w_i(S,\hat{S}^{(\epsilon)}_i)]\leq d_i+\epsilon$ for every $\epsilon>0$, $i=1,2$. Therefore, one can find a sequence $\epsilon_1, \epsilon_2, \cdots$ converging to zero such that
\begin{align*}
\lim\limits_{k\rightarrow\infty}p_{S,\hat{S}^{(\epsilon_k)}_1,\hat{S}^{(\epsilon_k)}_2}= p_{S,\hat{S}_1,\hat{S}_2}
\end{align*}
for some $p_{S,\hat{S}_1,\hat{S}_2}$ with $p_{S,\hat{S}_i}\in\mathcal{Q}_i(w_i,d_i)$, $i=1,2$. This completes the proof of the ``only if" part.
 \end{IEEEproof}

Source-channel separation is known to incur a performance loss for System $\Pi$ in general. However, it turns out that, for the following variant of System $\Pi$ (see Fig. \ref{fig:systemB}),  separate source-channel coding is in fact optimal.  This system (System $\tilde{\Pi}$) is the same as System $\Pi$ except for two differences.
\begin{enumerate}
\item The source is an i.i.d. vector process $\{(\tilde{S}_1(t),\tilde{S}_2(t))\}_{t=1}^\infty$ with marginal distribution $p_{\tilde{S}_1,\tilde{S}_2}$ over finite alphabet $\tilde{\mathcal{S}}_1\times\tilde{\mathcal{S}}_2$.

\item $\tilde{S}^m_{2}$ is available at receiver 1 and can be used together with $Y^n_{1}$ to construct $\hat{S}^m_{1}$.
\end{enumerate}

Let $\mathcal{P}_{\tilde{\mathcal{S}}_1\times\tilde{\mathcal{S}}_2\times\hat{\mathcal{S}}_1}(p_{\tilde{S}_1,\tilde{S}_2})$ denote the set of joint distributions over $\tilde{\mathcal{S}}_1\times\tilde{\mathcal{S}}_2\times\hat{\mathcal{S}}_1$ with the marginal distribution on $\tilde{\mathcal{S}}_1\times\tilde{\mathcal{S}}_2$ fixed to be $p_{\tilde{S}_1,\tilde{S}_2}$. Moreover, let $\mathcal{P}_{\tilde{\mathcal{S}}_2\times\hat{\mathcal{S}}_2}(p_{\tilde{S}_2})$ denote the set of joint distributions over $\tilde{\mathcal{S}}_2\times\hat{\mathcal{S}}_2$ with the marginal distribution on $\tilde{\mathcal{S}}_2$ fixed to be $p_{\tilde{S}_2}$.

\begin{definition}\label{def:systemB}
 Let $\tilde{\kappa}$ be a non-negative number,  $\tilde{\mathcal{Q}}_1$ be a non-empty compact subset of  $\mathcal{P}_{\tilde{\mathcal{S}}_1\times\tilde{\mathcal{S}}_2\times\hat{\mathcal{S}}_1}(p_{\tilde{S}_1,\tilde{S}_2})$, and  $\tilde{\mathcal{Q}}_2$ be a non-empty compact subset of  $\mathcal{P}_{\tilde{\mathcal{S}}_2\times\hat{\mathcal{S}}_2}(p_{\tilde{S}_2})$.  We say $(\tilde{\kappa},\tilde{\mathcal{Q}}_1,\tilde{\mathcal{Q}}_2)$ is achievable for System $\tilde{\Pi}$ if, for every $\epsilon>0$, there exist encoding function $f^{(m,n)}:\tilde{\mathcal{S}}^m_1\times\tilde{\mathcal{S}}^m_2\rightarrow\mathcal{X}^n$ as well as decoding functions $g^{(n,m)}_1:\mathcal{Y}^n_1\times\tilde{\mathcal{S}}^m_2\rightarrow\hat{\mathcal{S}}^m_1$ and $g^{(n,m)}_2:\mathcal{Y}^n_2\rightarrow\hat{\mathcal{S}}^m_2$ such that
\begin{align}
&\frac{n}{m}\leq\tilde{\kappa}+\epsilon,\label{eq:sysB1}\\
&\min\limits_{\tilde{q}_1\in\tilde{\mathcal{Q}}_1}\left\|\frac{1}{m}\sum\limits_{t=1}^mp_{\tilde{S}_1(t),\tilde{S}_2(t),\hat{S}_1(t)}-\tilde{q}_1\right\|\leq\epsilon,\label{eq:sysB2}\\
&\min\limits_{\tilde{q}_2\in\tilde{\mathcal{Q}}_2}\left\|\frac{1}{m}\sum\limits_{t=1}^mp_{\tilde{S}_2(t),\hat{S}_2(t)}-\tilde{q}_2\right\|\leq\epsilon.\label{eq:sysB3}
\end{align}
The set of all achievable $(\tilde{\kappa}, \tilde{\mathcal{Q}}_1,\tilde{\mathcal{Q}}_2)$ for System $\tilde{\Pi}$ is denoted by $\tilde{\Gamma}$.
\end{definition}

Remark: For the ease of subsequent applications, here we allow $f^{(m,n)}$, $g^{(n,m)}_1$, and $g^{(n,m)}_2$ to be non-deterministic functions as long as the Markov chains $(\tilde{S}^m_{1},\tilde{S}^m_{2})\leftrightarrow X^n\leftrightarrow (Y^n_{1},Y^n_{2})$, $\tilde{S}^m_{1}\leftrightarrow(Y^n_{1},\tilde{S}^m_{2})\leftrightarrow\hat{S}^m_{1}$, and $\tilde{S}^m_{2}\leftrightarrow Y^n_{2}\leftrightarrow\hat{S}^m_{2}$ are preserved. It will be clear that such a relaxation does not affect $\tilde{\Gamma}$.

 To discuss source-channel separation for System $\tilde{\Pi}$, we need to specify the source coding component and the channel coding component. It will be seen that the source coding part is the conventional lossy source coding scheme. The channel coding part is more involved and is described in the next section.

\begin{figure*}
\centering
\begin{tikzpicture}
\node [] (Source) {$M_1,M_2$};
\node [block, right of=Source , xshift=0.6em] (Enc) {transmitter\\ $f^{(n)}$};
\node [chblock, right of=Enc , xshift=1em] (Channel) {$p_{Y_{1},Y_{2}|X}$};\
\node [block, right of=Channel , xshift=1em, yshift=3.5em] (Dec1) {receiver 1\\ $g_1^{(n)}$};
\node [block, right of=Channel , xshift=1em, yshift=-3.5em] (Dec2) {receiver 2\\ $g_2^{(n)}$};
\node [right of=Dec1 , xshift=4.8em] (Rec1) {$\hat{M}_1$};
\node [right of=Dec2 , xshift=4.8em] (Rec2) {$\hat{M}_2$};
\path [line] (19.9em,3.55em)--node[xshift=0em,above] {$Y_{1}^n$}(22.7em,3.55em);
\path [line] (19.9em,-3.55em)--node[xshift=0em,above] {$Y_{1}^n$}(22.7em,-3.55em);
\path [line] (Source)--(Enc);
\path [line] (Enc) -- node[xshift=0em,above] {$X_1^n$} (Channel);
\path [line] (Dec1)--(Rec1);
\path [line] (Dec2)--(Rec2);
\end{tikzpicture}
\caption{Broadcast channel with two private messages}\label{fig:broadcast}
\end{figure*}
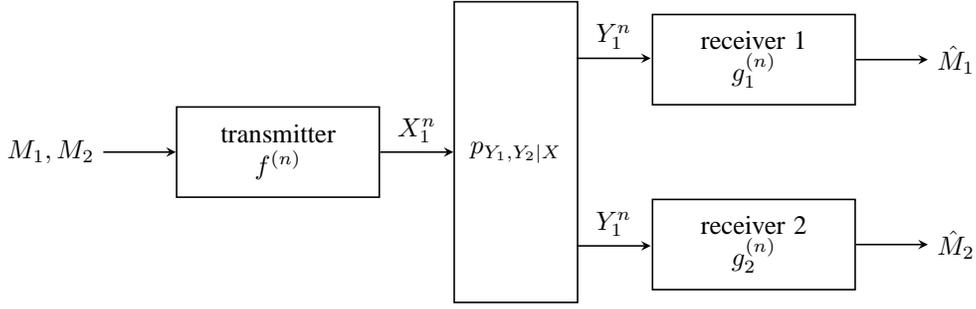

\begin{figure*}
\centering
\begin{tikzpicture}
\node [] (Source) {$M_1,M_2$};
\node [block, right of=Source , xshift=0.5em] (Enc) {transmitter\\ $f^{(n)}$};
\node [chblock, right of=Enc , xshift=1em] (Channel) {$p_{Y_{1},Y_{2}|X}$};\
\node [block, right of=Channel , xshift=1em, yshift=3.5em] (Dec1) {receiver 1\\ $g_1^{(n)}$};
\node [above of=Dec1, yshift=2.5em] (Side) {$M_2$};
\node [block, right of=Channel , xshift=1em, yshift=-3.5em] (Dec2) {receiver 2\\ $g_2^{(n)}$};
\node [right of=Dec1 , xshift=4.8em] (Rec1) {$\hat{M}_1$};
\node [right of=Dec2 , xshift=4.8em] (Rec2) {$\hat{M}_2$};
\path [line] (19.8em,3.55em)--node[xshift=0em,above] {$Y_{1}^n$}(22.65em,3.55em);
\path [line] (19.8em,-3.55em)--node[xshift=0em,above] {$Y_{2}^n$}(22.65em,-3.55em);
\path [line] (Source)--(Enc);
\path [line] (Enc) -- node[xshift=0em,above] {$X_1^n$} (Channel);
\path [line] (Dec1)--(Rec1);
\path [line] (Dec2)--(Rec2);
\path [line] (Side)--(Dec1);
\end{tikzpicture}
\caption{Broadcast channel with receiver side information}\label{fig:broadside}
\end{figure*}
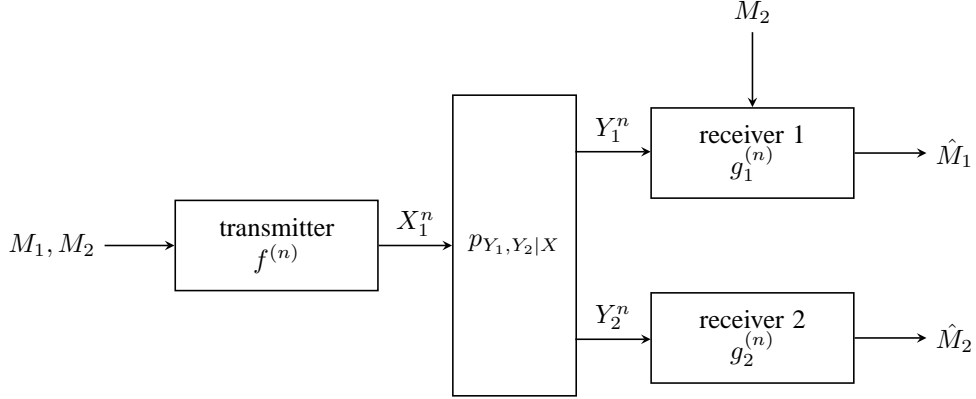

\section{Broadcast Channels
with Receiver Side Information}\label{sec:broadcast}

\subsection{Definitions}

Let $p_{Y_1,Y_2|X}$ be a discrete memoryless broadcast channel  with input alphabet $\mathcal{X}$ and output alphabets $\mathcal{Y}_i$, $i=1,2$. A length-$n$ coding scheme (see Fig. \ref{fig:broadcast}) for $p_{Y_1,Y_2|X}$ consists of
\begin{itemize}
\item two private messages $M_1$ and $M_2$, where $(M_1,M_2)$ is uniformly distributed over $\mathcal{M}_1\times\mathcal{M}_2$,

\item an encoding function $f^{(n)}:\mathcal{M}_1\times\mathcal{M}_2\rightarrow\mathcal{X}^n$ that maps $(M_1,M_2)$ to a channel input block $X^n$,

\item two decoding functions $g^{(n)}_i:\mathcal{Y}^n_i\rightarrow\mathcal{M}_i$, $i=1,2$, where $g^{(n)}_i$ maps the channel output block at receiver $i$, i.e., $Y^n_{i}$, to $\hat{M}_i$, $i=1,2$.
\end{itemize}

\begin{definition}
A rate pair $(R_1,R_2)\in\mathbb{R}^2_+$ is said to be achievable for broadcast channel $p_{Y_1,Y_2|X}$ if there exists a sequence of encoding functions $f^{(n)}:\mathcal{M}_1\times\mathcal{M}_2\rightarrow\mathcal{X}^n$ with $\frac{1}{n}\log|\mathcal{M}_i|\geq R_i$, $i=1,2$, and decoding functions $g^{(n)}_i:\mathcal{Y}^n_i\rightarrow\mathcal{M}_i$, $i=1,2$, such that
\begin{align*}
\lim\limits_{n\rightarrow\infty}\mbox{Pr}\{(\hat{M}_1,\hat{M}_2)\neq(M_1,M_2)\}=0.
\end{align*}
The private-message capacity region $\mathcal{C}(p_{Y_1,Y_2|X})$ is the closure of the set of all achievable $(R_1,R_2)$ for broadcast channel $p_{Y_1,Y_2|X}$.
\end{definition}

A computable characterization of $\mathcal{C}(p_{Y_1,Y_2|X})$  is still largely unknown. Interestingly, the problem becomes significantly simpler if message $M_2$ is available at receiver 1 or message $M_1$ is available at receiver 2; in fact, this is the setting that is most relevant to the present work.
Specifically, consider the scenario where two private messages $M_1$ and $M_2$ need to be sent over broadcast channel $p_{Y_1,Y_2|X}$ to receiver 1 and receiver 2, respectively, and $M_2$ is available at receiver 1. In this case,  a length-$n$ coding scheme (see Fig. \ref{fig:broadside}) consists of
\begin{itemize}
\item two private messages $M_i$, $i=1,2$, where $(M_1,M_2)$ is uniformly distributed over $\mathcal{M}_1\times\mathcal{M}_2$,

\item an encoding function $f^{(n)}:\mathcal{M}_1\times\mathcal{M}_2\rightarrow\mathcal{X}^n$ that maps $(M_1,M_2)$ to a channel input block $X^n$,

\item two decoding functions $g^{(n)}_1:\mathcal{Y}^n_1\times\mathcal{M}_2\rightarrow\mathcal{M}_1$ and $g^{(n)}_2:\mathcal{Y}^n_2\rightarrow\mathcal{M}_2$, where $g^{(n)}_1$ maps $(Y^n_{1},M_2)$ to $\hat{M}_{1}$,  and $g^{(n)}_2$ maps $Y^n_{2}$ to $\hat{M}_{2}$.
\end{itemize}

\begin{definition}\label{def:broadcastside}
A rate pair $(R_1,R_2)$ is said to be achievable for broadcast channel $p_{Y_1,Y_2|X}$ with message $M_2$ available at receiver 1 if there exists a sequence of encoding functions $f^{(n)}:\mathcal{M}_1\times\mathcal{M}_2\rightarrow\mathcal{X}^n$ with $\frac{1}{n}\log|\mathcal{M}_i|\geq R_i$, $i=1,2$, as well as decoding functions $g_1^{(n)}:\mathcal{Y}^n_1\times\mathcal{M}_2\rightarrow\mathcal{M}_1$ and $g_2^{(n)}:\mathcal{Y}^n_2\rightarrow\mathcal{M}_2$ such that
\begin{align*}
\lim\limits_{n\rightarrow\infty}\mbox{Pr}\{(\hat{M}_{1},\hat{M}_2)\neq(M_1,M_2)\}=0.
\end{align*}
The capacity region $\mathcal{C}_{1}(p_{Y_1,Y_2|X})$ is the closure of the set of all such achievable $(R_1,R_2)$. The capacity region $\mathcal{C}_{2}(p_{Y_1,Y_2|X})$ for broadcast channel $p_{Y_1,Y_2|X}$ with message $M_1$ available at receiver 2 can be defined in an analogous manner.
\end{definition}

\subsection{Capacity Results}

It is known \cite[Theorem 3]{KS07} that $\mathcal{C}_{1}(p_{Y_1,Y_2|X})$ is given by the set of $(R_1,R_2)\in\mathbb{R}^2_+$ satisfying
\begin{align}
&R_1\leq I(X;Y_1),\label{eq:redundant}\\
&R_2\leq I(V;Y_2),\label{eq:nr1}\\
&R_1+R_2\leq I(X;Y_1|V)+I(V;Y_2)\label{eq:nr2}
\end{align}
for some $p_{V,X,Y_1,Y_2}=p_{V,X}p_{Y_1,Y_2|X}$; moreover, it suffices to assume that $|\mathcal{V}|\leq|\mathcal{X}|+1$. By symmetry,
$\mathcal{C}_{2}(p_{Y_1,Y_2|X})$ is given by the set of $(R_1,R_2)\in\mathbb{R}^2_+$ satisfying
\begin{align}
&R_1\leq I(V;Y_1),\label{eq:sw1}\\
&R_2\leq I(X;Y_2),\label{eq:se2}\\
&R_1+R_2\leq I(V;Y_1)+ I(X;Y_2|V)\label{eq:sw3}
\end{align}
for some $p_{V,X,Y_1,Y_2}=p_{V,X}p_{Y_1,Y_2|X}$; again, it suffices to assume that $|\mathcal{V}|\leq|\mathcal{X}|+1$.

A class of distributions $\mathcal{P}$ on the
input alphabet $\mathcal{X}$ is said to be a sufficient class of distributions \cite[Definition 1]{Nair10} for broadcast channel $p_{Y_1,Y_2|X}$ if, for any $p_{V_1,V_2,X,Y_1,Y_2}=p_{V_1,V_2,X}p_{Y_1,Y_2|X}$, there exists $p_{\tilde{V}_1,\tilde{V}_2,\tilde{X},\tilde{Y}_1,\tilde{Y}_2}=p_{\tilde{V}_1,\tilde{V}_2,\tilde{X}}p_{\tilde{Y}_1,\tilde{Y}_2|\tilde{X}}$ with $p_{\tilde{X}}\in\mathcal{P}$ and $p_{\tilde{Y}_1,\tilde{Y}_2|\tilde{X}}=p_{Y_1,Y_2|X}$ such that\footnote{Setting $V_1=X$, one can readily verify that $I(X;Y_1)=I(V_1;Y_1)\leq I(\tilde{V}_1;\tilde{Y}_1)\leq I(\tilde{X};\tilde{Y}_1)$. Similarly, one can obtain $I(X;Y_2)\leq I(\tilde{X};\tilde{Y}_2)$ by setting $V_2=X$.}
\begin{align*}
&I(V_1;Y_1)\leq I(\tilde{V}_1;\tilde{Y}_1),\\
&I(V_2;Y_2)\leq I(\tilde{V}_2;\tilde{Y}_2),\\
&I(V_1;Y_1)+I(X;Y_2|V_1)\leq I(\tilde{V}_1;\tilde{Y}_1)+I(\tilde{X};\tilde{Y}_2|\tilde{V}_1),\\
&I(X;Y_1|V_2)+I(V_2;Y_2)\leq I(\tilde{X};\tilde{Y}_1|\tilde{V}_2)+I(\tilde{V}_2;\tilde{Y}_2).
\end{align*}
For broadcast channel $p_{Y_1,Y_2|X}$, we say that $p_{Y_1|X}$ is essentially less noisy than $p_{Y_2|X}$ if there exists a sufficient class of distributions $\mathcal{P}$ such that
$I(V;Y_1)\geq I(V;Y_2)$ for any $p_{V,X,Y_1,Y_2}=p_{V,X}p_{Y_1,Y_2|X}$ with $p_X\in\mathcal{P}$ \cite[Definition 2]{Nair10}, and simply say that $p_{Y_1|X}$ is less noisy than $p_{Y_2|X}$ if $\mathcal{P}$ can be chosen to be the set of all distributions on $\mathcal{X}$; similarly, we say that $p_{Y_1|X}$ is essentially more capable than $p_{Y_2|X}$ if there exists a sufficient class of distributions $\mathcal{P}$ such that
$I(X;Y_1|V)\geq I(X;Y_2|V)$ for any $p_{V,X,Y_1,Y_2}=p_{V,X}p_{Y_1,Y_2|X}$ with $p_X\in\mathcal{P}$ \cite[Definition 3]{Nair10}, and simply say that $p_{Y_1|X}$ is more capable than $p_{Y_2|X}$ if $\mathcal{P}$ can be chosen to be the set of all distributions on $\mathcal{X}$. It is known that ``less noisy" (``more capable") implies ``essentially less noisy" (``essentially more capable"), and ``less noisy" implies ``more capable", but the converses are not true in general.


\begin{proposition}\label{prop:lessnoisy}
If $p_{Y_1|X}$ is essentially less noisy than $p_{Y_2|X}$, then $\mathcal{C}_{1}(p_{Y_1,Y_2|X})=\mathcal{C}(p_{Y_1,Y_2|X})$.
\end{proposition}
\begin{IEEEproof}
To compute $\mathcal{C}_{1}(p_{Y_1,Y_2|X})$ defined by (\ref{eq:redundant})-(\ref{eq:nr2}), it suffices to consider those $p_X$ in a sufficient class $\mathcal{P}$. It is easy to see that
\begin{align}
 I(X;Y_1|V)+I(V;Y_2)&\leq  I(X;Y_1|V)+I(V;Y_1)\label{eq:lessnoisy}\\
 &=I(X;Y_1)\nonumber
\end{align}
for any $p_{V,X,Y_1,Y_2}=p_{V,X}p_{Y_1,Y_2|X}$ with $p_X\in\mathcal{P}$, where (\ref{eq:lessnoisy}) is due to the fact that $p_{Y_1|X}$ is essentially less noisy than $p_{Y_2|X}$. Therefore, (\ref{eq:redundant}) is redundant if $p_X$ is restricted to $\mathcal{P}$.
Note that the rate region defined by (\ref{eq:nr1}) and (\ref{eq:nr2}) for $p_{V,X,Y_1,Y_2}=p_{V,X}p_{Y_1,Y_2|X}$ with $p_X\in\mathcal{P}$ is exactly $\mathcal{C}(p_{Y_1,Y_2|X})$ \cite[Theorem 1]{Nair10}. This completes the proof of Proposition \ref{prop:lessnoisy}.
\end{IEEEproof}

\begin{proposition}\label{prop:morecapable}
If  $p_{Y_1|X}$ is essentially more capable than $p_{Y_2|X}$, then $\mathcal{C}_{2}(p_{Y_1,Y_2|X})$ is given by the set of $(R_1,R_2)\in\mathbb{R}^2_+$ satisfying
\begin{align*}
&R_2\leq I(X;Y_2),\\
& R_1+R_2\leq I(X;Y_1)
\end{align*}
for some  $p_{X,Y_1,Y_2}=p_Xp_{Y_1,Y_2|X}$.
\end{proposition}
\begin{IEEEproof}
To compute $\mathcal{C}_{2}(p_{Y_1,Y_2|X})$ defined by (\ref{eq:sw1})-(\ref{eq:sw3}), it suffices to consider those $p_X$ in a sufficiently class $\mathcal{P}$. Note that
\begin{align}
I(V;Y_1)+ I(X;Y_2|V)&\leq I(V;Y_1)+ I(X;Y_1|V)\label{eq:mc}\\
&=I(X;Y_1)\nonumber
\end{align}
for any $p_{V,X,Y_1,Y_2}=p_{V,X}p_{Y_1,Y_2|X}$ with $p_X\in\mathcal{P}$, where (\ref{eq:mc}) is due to the fact that $p_{Y_1|X}$ is essentially more capable than $p_{Y_2|X}$. Therefore,  given $p_{X}\in\mathcal{P}$, the right-hand side of inequality (\ref{eq:sw3}) attains its maximum value $I(X;Y_1)$ when $V=X$. Clearly, given $p_{X}$, the right-hand side of inequality (\ref{eq:sw1}) also attains its maximum value $I(X;Y_1)$ when $V=X$. As a consequence,
$\mathcal{C}_{2}(p_{Y_1,Y_2|X})$ can be expressed as the set of $(R_1,R_2)\in\mathbb{R}^2_+$ satisfying
\begin{align*}
&R_2\leq I(X;Y_2),\\
& R_1+R_2\leq I(X;Y_1)
\end{align*}
for some  $p_{X,Y_1,Y_2}=p_Xp_{Y_1,Y_2|X}$ with $p_X\in\mathcal{P}$.  Removing the redundant constraint $p_X\in\mathcal{P}$ completes the proof of Proposition \ref{prop:morecapable}.
\end{IEEEproof}

\subsection{Examples}\label{sec:IIIexample}

 Consider a broadcast channel $p_{Y_1,Y_2|X}$ with $\mathcal{X}=\mathcal{Y}_1=\mathcal{Y}_2=\{0,1\}$, where $p_{Y_i|X}$ is a binary symmetric channel with crossover probability $p_i$, $i=1,2$; such a channel will be denoted by  $\mbox{BS-BC}(p_1,p_2)$. Without loss of generality, we shall assume $0\leq p_1\leq p_2\leq\frac{1}{2}$. It is well known that $\mathcal{C}(\mbox{BS}(p_1,p_2))$ is given by the set of $(R_1,R_2)\in\mathbb{R}^2_+$ satisfying
\begin{align*}
&R_1\leq H_b(\alpha*p_1)-H_b(p_1),\\
&R_2\leq 1-H_b(\alpha*p_2)
\end{align*}
for some $\alpha\in[0,\frac{1}{2}]$. Next consider a broadcast channel $p_{Y_1,Y_2|X}$ with $\mathcal{X}=\{0,1\}$ and $\mathcal{Y}_i=\{0,1,e\}$, $i=1,2$, where $p_{Y_i|X}$ is a binary erasure channel with erasure probability $\epsilon_i$, $i=1,2$; such a channel will be denoted by $\mbox{BE-BC}(\epsilon_1,\epsilon_2)$. Without loss of generality, we shall assume $0\leq\epsilon_1\leq\epsilon_2\leq 1$. It is well known that $\mathcal{C}(\mbox{BE-BC}(\epsilon_1,\epsilon_2))$ is given by the set of $(R_1,R_2)\in\mathbb{R}^2_+$ satisfying
\begin{align}
&R_1\leq\beta(1-\epsilon_1),\label{eq:bebcexp1}\\
&R_2\leq(1-\beta)(1-\epsilon_2)\label{eq:bebcexp2}
\end{align}
for some $\beta\in[0,1]$.

The following results are simple consequences of Proposition  \ref{prop:lessnoisy} and Proposition \ref{prop:morecapable}.

\begin{proposition}\label{prop:bsbc}
For $\mbox{BS-BC}(p_1,p_2)$ with $0\leq p_1\leq p_2\leq\frac{1}{2}$,
\begin{align*}
&\mathcal{C}_{1}(\mbox{BS-BC}(p_1,p_2))=\mathcal{C}(\mbox{BS-BC}(p_1,p_2)),\\
&\mathcal{C}_{2}(\mbox{BS-BC}(p_1,p_2))=\{(R_1,R_2)\in\mathbb{R}^2_+: R_2\leq 1-H_b(p_2),\\
&\hspace{1.98in}R_1+R_2\leq 1-H_b(p_1)\}.
\end{align*}
\end{proposition}

\begin{proposition}\label{prop:bebc}
For $\mbox{BE-BC}(\epsilon_1,\epsilon_2)$ with $0\leq\epsilon_1\leq\epsilon_2\leq 1$,
\begin{align*}
&\mathcal{C}_{1}(\mbox{BE-BC}(\epsilon_1,\epsilon_2))=\mathcal{C}(\mbox{BE-BC}(\epsilon_1,\epsilon_2)),\\
&\mathcal{C}_{2}(\mbox{BE-BC}(\epsilon_1,\epsilon_2))=\{(R_1,R_2)\in\mathbb{R}^2_+:R_2\leq 1-\epsilon_2, \\
&\hspace{1.96in}R_1+R_2\leq 1-\epsilon_1\}.
\end{align*}
\end{proposition}

Now consider a broadcast channel $p_{Y_1,Y_2|X}$ with $\mathcal{X}=\mathcal{Y}_1=\{0,1\}$ and $\mathcal{Y}_2=\{0,1,e\}$,  where $p_{Y_1|X}$ is a binary symmetric channel with crossover probability $p$, and $p_{Y_2|X}$ is a binary erasure channel with erasure probability $\epsilon$; such a channel will be denoted by $\mbox{BSC}(p)\&\mbox{BEC}(\epsilon)$. Without loss of generality, we shall assume $p\in[0,\frac{1}{2}]$ and $\epsilon\in[0,1]$.
One can obtain the following explicit characterization of $\mathcal{C}(\mbox{BSC}(p)\&\mbox{BEC}(\epsilon))$ \cite[Theorem 4]{Nair10}.
\begin{enumerate}
\item $\epsilon\in[0,4p(1-p)]$: $\mathcal{C}(\mbox{BSC}(p)\&\mbox{BEC}(\epsilon))$ is given by the set of $(R_1,R_2)\in\mathbb{R}^2_+$ satisfying
\begin{align*}
&R_1\leq 1-H_b(\alpha*p),\\
&R_2\leq (1-\epsilon)H_b(\alpha)
\end{align*}
for some $\alpha\in[0,\frac{1}{2}]$.

\item $\epsilon\in(4p(1-p),H_b(p))$: $\mathcal{C}(\mbox{BSC}(p)\&\mbox{BEC}(\epsilon))$ is given by the set of $(R_1,R_2)\in\mathbb{R}^2_+$ satisfying
\begin{align*}
&R_1\leq 1-H_b(\alpha*p),\\
&R_2\leq (1-\epsilon)H_b(\alpha)
\end{align*}
for some $\alpha\in[0,\hat{\alpha}]$, or
\begin{align*}
&R_1\leq 1-H_b(\alpha*p),\\
&R_2\leq H_b(\alpha*p)-\epsilon
\end{align*}
for some $\alpha\in(\hat{\alpha},\frac{1}{2}]$,
where $\hat{\alpha}$ is the unique number in $(0,\frac{1}{2})$ satisfying
\begin{align*}
1-H_b(\hat{\alpha}*p)+(1-\epsilon)H_b(\hat{\alpha})=1-\epsilon.
\end{align*}

\item $\epsilon\in[H_b(p),1]$: $\mathcal{C}(\mbox{BSC}(p)\&\mbox{BEC}(\epsilon))$ is given by the set of $(R_1,R_2)\in\mathbb{R}^2_+$ satisfying
\begin{align*}
&R_1\leq \beta[1-H_b(p)],\\
&R_2\leq (1-\beta)(1-\epsilon)
\end{align*}
for some $\beta\in[0,1]$.
\end{enumerate}

\begin{proposition}\label{prop:becbsc21}
$\mathcal{C}_{1}(\mbox{BSC}(p)\&\mbox{BEC}(\epsilon))$ has the following explicit characterization.
\begin{enumerate}
\item $\epsilon\in[0,H_b(p)]$:
\begin{align*}
&\mathcal{C}_{1}(\mbox{BSC}(p)\&\mbox{BEC}(\epsilon))=\{(R_1,R_2)\in\mathbb{R}^2_+: \\
&\hspace{1.0in}R_1\leq 1-H_b(p), R_1+R_2\leq 1-\epsilon\}.
\end{align*}

\item $\epsilon\in(H_b(p),1]$: 
\begin{align*}
\mathcal{C}_{1}(\mbox{BSC}(p)\&\mbox{BEC}(\epsilon))=\mathcal{C}(\mbox{BSC}(p)\&\mbox{BEC}(\epsilon)).
\end{align*}
\end{enumerate}
\end{proposition}
\begin{IEEEproof}
According to  \cite[Theorem 3]{Nair10}, $\mbox{BEC}(\epsilon)$ is more capable than $\mbox{BSC}(p)$ when $\epsilon\in[0,H_b(p)]$. Therefore, one can readily prove Part 1) by invoking Proposition \ref{prop:morecapable} as well as the fact that $I(X;Y_1)$ and $I(X;Y_2)$ are simultaneously maximized when $p_X(0)=p_X(1)=\frac{1}{2}$.
Part 2) follows from Proposition \ref{prop:lessnoisy} and the fact that $\mbox{BSC}(p)$ is essentially less noisy than $\mbox{BEC}(\epsilon)$ when $\epsilon\in(H_b(p),1]$ \cite[Theorem 3]{Nair10}.
\end{IEEEproof}

\begin{proposition}\label{prop:becbsc12}
$\mathcal{C}_{2}(\mbox{BSC}(p)\&\mbox{BEC}(\epsilon))$ has the following explicit characterization.
\begin{enumerate}
\item $\epsilon\in[0, 4p(1-p)]$: 
\begin{align*}
\mathcal{C}_{2}(\mbox{BSC}(p)\&\mbox{BEC}(\epsilon))=\mathcal{C}(\mbox{BSC}(p)\&\mbox{BEC}(\epsilon)).
\end{align*}

\item $\epsilon\in(4p(1-p),1)$ and $p\neq 0$: $\mathcal{C}_{2}(\mbox{BSC}(p)\&\mbox{BEC}(\epsilon))$ is given by the set of $(R_1,R_2)\in\mathbb{R}^2_+$ satisfying
\begin{align*}
&R_1\leq 1-H_b(\alpha*p),\\
&R_2\leq (1-\epsilon)H_b(\alpha)
\end{align*}
for some $\alpha\in[0,\tilde{\alpha}]$, or
\begin{align*}
&R_1\leq 1-H_b(\tilde{\alpha}*p),\\
&R_2\leq 1-\epsilon,\\
&R_1+R_2\leq 1-H_b(\tilde{\alpha}*p)+(1-\epsilon)H_b(\tilde{\alpha})
\end{align*}
for some $\alpha\in(\tilde{\alpha},\frac{1}{2}]$, where $\tilde{\alpha}$ is the unique number in $(0,\frac{1}{2})$ satisfying
\begin{align*}
(1-2p)\log\Big(\frac{1-\tilde{\alpha}*p}{\tilde{\alpha}*p}\Big)=(1-\epsilon)\log\Big(\frac{1-\tilde{\alpha}}{\tilde{\alpha}}\Big).
\end{align*}

\item $\epsilon=1$ or $p=0$:
\begin{align*}
&\mathcal{C}_{2}(\mbox{BSC}(p)\&\mbox{BEC}(\epsilon))=\{(R_1,R_2)\in\mathbb{R}^2_+:\\
&\hspace{1.0in}R_2\leq 1-\epsilon, R_1+R_2\leq 1-H_b(p)\}.
\end{align*}
\end{enumerate}
\end{proposition}
\begin{IEEEproof}
Part 1) follows from Proposition \ref{prop:lessnoisy} and the fact that $\mbox{BEC}(\epsilon)$ is less noisy than $\mbox{BSC}(p)$ when $\epsilon\in[0, 4p(1-p)]$ \cite[Theorem 3]{Nair10}. Part 3) is trivial. For Part 2), one can readily show that $\mathcal{C}_{2}(\mbox{BSC}(p)\&\mbox{BEC}(\epsilon))$ is given by the set of $(R_1,R_2)\in\mathbb{R}^2_+$ satisfying
\begin{align*}
&R_1\leq 1-H_b(\alpha*p),\\
&R_2\leq 1-\epsilon,\\
&R_1+R_2\leq 1-H_b(\alpha*p)+(1-\epsilon)H_b(\alpha)
\end{align*}
for some $\alpha\in[0,\frac{1}{2}]$ by following the proof of \cite[Claim 2 and Claim 3]{Nair10}. In light of \cite[Lemma 6]{SSC13}, when $\epsilon\in(4p(1-p),1)$ and $p\neq 0$, the following optimization problem
\begin{align*}
\max\limits_{\alpha\in[0,\frac{1}{2}]}1-H_b(\alpha*p)+(1-\epsilon)H_b(\alpha)
\end{align*}
has a unique maximizer at $\alpha=\tilde{\alpha}$. This completes the proof of Proposition \ref{prop:becbsc12}.
\end{IEEEproof}

\begin{figure}
       \centering   \includegraphics[width = 3.1in]{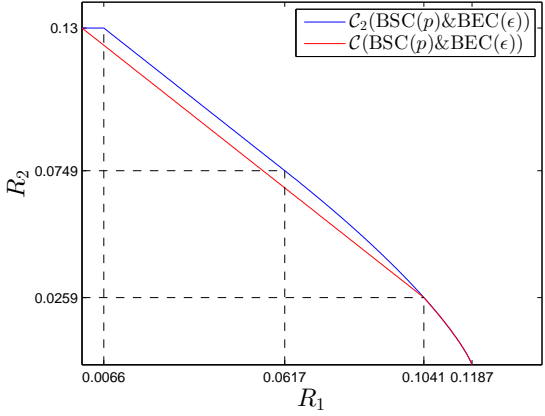}
       \caption{$\mathcal{C}_{2}(\mbox{BSC}(p)\&\mbox{BEC}(\epsilon))$ vs. $\mathcal{C}(\mbox{BEC}(\epsilon)\&\mbox{BSC}(p))$ with $p=0.3$ and $\epsilon=0.87$}\label{fig:becbsc1}
\end{figure}

\begin{figure}
       \centering   \includegraphics[width = 3.1in]{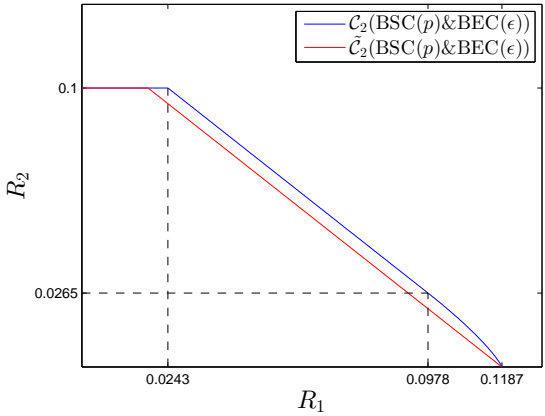}
       \caption{$\mathcal{C}_{2}(\mbox{BSC}(p)\&\mbox{BEC}(\epsilon))$ vs. $\tilde{\mathcal{C}}_{2}(\mbox{BSC}(p)\&\mbox{BEC}(\epsilon))\triangleq\{(R_1,R_2)\in\mathbb{R}^2_+:R_2\leq 1-\epsilon,R_1+R_2\leq 1-H_b(p)\}$ with $p=0.3$ and $\epsilon=0.9$}\label{fig:becbsc2}
\end{figure}

Remark: It might be tempting to conjecture that Proposition \ref{prop:lessnoisy} continues to hold if ``essentially less noisy" is replaced by ``essentially more capable". However, this conjecture turns out to be false. Indeed, for $\mbox{BSC}(p)\&\mbox{BEC}(\epsilon)$, it is known \cite[Theorem 3]{Nair10} that $\mbox{BEC}(\epsilon)$ is more capable (but not less noisy) than $\mbox{BSC}(p)$ when $\epsilon\in(4p(1-p),H_b(p)]$, yet Part 2) of Proposition \ref{prop:becbsc12} indicates that in this case $\mathcal{C}_{2}(\mbox{BSC}(p)\&\mbox{BEC}(\epsilon))$ is strictly larger than $\mathcal{C}(\mbox{BSC}(p)\&\mbox{BEC}(\epsilon))$ (see Fig. \ref{fig:becbsc1}). Analogously, Proposition \ref{prop:morecapable} is not true in general if ``essentially more capable" is replaced by ``essentially less noisy". For example, according to \cite[Theorem 3]{Nair10} , $\mbox{BSC}(p)$ is essentially less noisy than $\mbox{BEC}(\epsilon)$ when $\epsilon\in[H_b(p),1)$ and $p\neq 0$, but Part 2) of Proposition \ref{prop:becbsc12} shows that in this case $\mathcal{C}_{2}(\mbox{BSC}(p)\&\mbox{BEC}(\epsilon))$ is strictly larger than $\{(R_1,R_2)\in\mathbb{R}^2_+:R_2\leq 1-\epsilon,R_1+R_2\leq 1-H_b(p)\}$ (see Fig. \ref{fig:becbsc2}).

Finally consider the case where $p_{Y_1,Y_2|X}$ is a scalar Gaussian broadcast channel with power constraint $P$ and noise variances $N_1$ and $N_2$ ($0<N_1\leq N_2$); such a channel will be denoted by  $\mbox{G-BC}(P, N_1,N_2)$. It is well known that $\mathcal{C}(\mbox{G-BC}(P, N_1,N_2))$ is  given by the set of $(R_1,R_2)\in\mathbb{R}^2_+$ satisfying
\begin{align*}
&R_1\leq\frac{1}{2}\log\Big(\frac{\beta P+N_1}{N_1}\Big),\\
&R_2\leq\frac{1}{2}\log\Big(\frac{P+N_2}{\beta P+N_2}\Big)
\end{align*}
for some $\beta\in[0,1]$. One can readily prove the following result by adapting Proposition \ref{prop:lessnoisy} and Proposition \ref{prop:morecapable} to this channel model.
\begin{proposition}\label{prop:gbc}
For $\mbox{G-BC}(P,N_1,N_2)$ with $0<N_1\leq N_2$,
\begin{align*}
&\mathcal{C}_{1}(\mbox{G-BC}(P,N_1,N_2))=\mathcal{C}(\mbox{G-BC}(P,N_1,N_2)),\\
&\mathcal{C}_{2}(\mbox{G-BC}(P,N_1,N_2))=\Big\{(R_1,R_2)\in\mathbb{R}^2_+: \\
&\hspace{0.3in}R_2\leq\frac{1}{2}\log\Big(\frac{P+N_2}{N_2}\Big),R_1+R_2\leq\frac{1}{2}\log\Big(\frac{P+N_1}{N_1}\Big)\Big\}.
\end{align*}
\end{proposition}

\section{Optimality of Source-Channel Separation for System $\tilde{\Pi}$}\label{sec:separation}

Now we are in a position to state the following source-channel separation theorem, which shows that a separation-based scheme that consists of lossy source coding and broadcast channel coding (see Fig. \ref{fig:broadside} and the associated description) is optimal for System $\tilde{\Pi}$. This result can be viewed as an extension of \cite[Lemma 3]{GT13} from degraded broadcast channels to general broadcast channels.

\begin{theorem}\label{thm:systemB}
$(\tilde{\kappa}, \tilde{\mathcal{Q}}_1,\tilde{\mathcal{Q}}_2)\in\tilde{\Gamma}$ if and only if $(R_{\tilde{S}_1|\tilde{S}_2}(\tilde{\mathcal{Q}}_1),R_{\tilde{S}_2}(\tilde{\mathcal{Q}}_2))\in\tilde{\kappa}\mathcal{C}_{1}(p_{Y_1,Y_2|X})$, where
\begin{align*}
&R_{\tilde{S}_1|\tilde{S}_2}(\tilde{\mathcal{Q}}_1)=\min\limits_{p_{\tilde{S}_1,\tilde{S}_2,\hat{S}_1}\in\tilde{\mathcal{Q}}_1}I(\tilde{S}_1;\hat{S}_1|\tilde{S}_2),\\
&R_{\tilde{S}_2}(\tilde{\mathcal{Q}}_2)=\min\limits_{p_{\tilde{S}_2,\hat{S}_2}\in\tilde{\mathcal{Q}}_2}I(\tilde{S}_2;\hat{S}_2).
\end{align*}
\end{theorem}
\begin{IEEEproof}
The proof of the ``if" part hinges on a separation-based scheme. We shall only give a sketch here since the argument only involves standard techniques. Let $\hat{S}_1$ be jointly distributed with  $(\tilde{S}_1,\tilde{S}_2)$ such that  $p_{\tilde{S}_1,\tilde{S}_2,\hat{S}_1}\in\tilde{\mathcal{Q}}_1$ and $I(\tilde{S}_1;\hat{S}_1|\tilde{S}_2)=R_{\tilde{S}_1|\tilde{S}_2}(\tilde{\mathcal{Q}}_1)$. Let $\hat{S}_2$ be jointly distributed with $\tilde{S}_2$ such that $p_{\tilde{S}_2,\hat{S}_2}\in\tilde{\mathcal{Q}}_2$ and $I(\tilde{S}_2;\hat{S}_2)=R_{\tilde{S}_2}(\tilde{\mathcal{Q}}_2)$.
By the functional representation lemma \cite[p. 626]{EGK11} (see also \cite[Lemma 1]{WCZCP11}), we can find a random variable $W$ of cardinality $|\mathcal{W}|\leq|\tilde{\mathcal{S}}_2|(|\hat{\mathcal{S}}_1|-1)+1$ with the following properties:
\begin{itemize}
\item $W$ is independent of $\tilde{S}_2$;

\item $\hat{S}_1=\psi(\tilde{S}_2,W)$ for some deterministic function $\psi:\tilde{\mathcal{S}}_2\times\mathcal{W}\rightarrow\hat{\mathcal{S}}_1$;

\item $\tilde{S}_1\leftrightarrow(\tilde{S}_2,\hat{S}_1)\leftrightarrow W$ form a Markov chain.
\end{itemize}
It is easy to see that
\begin{align*}
I(\tilde{S}_1;\hat{S}_1|\tilde{S}_2)&=I(\tilde{S}_1;W|\tilde{S}_2)\\
&=I(\tilde{S}_1,\tilde{S}_2;W).
\end{align*}
For any $\delta>0$, let $R_1=(1+\delta)I(\tilde{S}_1;\hat{S}_1|\tilde{S}_2)$ and $R_2=(1+\delta)I(\tilde{S}_2;\hat{S}_2)$. We independently generate $2^{mR_1}$ codewords $W^m(m_1)$, $m_1=1,\cdots,2^{mR_1}$, each according to $\prod_{t=1}^mp_{W}$, and independently generate $2^{mR_2}$ codewords $\hat{S}^m_{2}(m_2)$, $m_2=1,\cdots,2^{mR_2}$, each according to $\prod_{t=1}^mp_{\hat{S}_2}$. Codebooks $\{W^m(m_1)\}_{m_1=1}^{2^{mR_1}}$ and $\{\hat{S}^m_{2}(m_2)\}_{m_2=1}^{2^{mR_2}}$ are revealed to the transmitter and the receivers.
It can be shown that, given $(\tilde{S}^m_{1},\tilde{S}^m_{2})$, with high probability one can find an index $M_1$ such that $(\tilde{S}^m_{1},\tilde{S}^m_{2},W^m(M_1))$ are jointly typical with respect to $p_{\tilde{S}_1,\tilde{S}_2,W}$ when $m$ is large enough (see \cite{EGK11} for the definition of typical sequences and the related properties). Similarly, given $\tilde{S}^m_{2}$, with high probability one can find an index $M_2$ such that $(\tilde{S}^m_{2},\hat{S}^m_{2}(M_2))$ are jointly typical with respect to $p_{\tilde{S}_2,\hat{S}_2}$. If there is more than one such $M_1$ (or $M_2$), we choose the smallest index among them; if no such $M_1$ (or $M_2$) exists, we set $M_1=1$ (or $M_2=1$). Now a length-$n$ coding scheme is used to send messages $M_1$ and $M_2$ over broadcast channel $p_{Y_1,Y_2|X}$ to receiver 1 and receiver 2, respectively. Given $\tilde{S}^m_{2}$, receiver 1 can recover $M_2$ and use it together with $Y^n_{1}$ to produce an estimate $\hat{M}_1$. Receiver 2 can use $Y^n_{2}$ to produce an estimate $\hat{M}_2$. We assume that this length-$n$ coding scheme is good in the sense that $M_i=\hat{M}_i$, $i=1,2$, with high probability.
Note that the existence of such a good length-$n$ coding scheme  is guaranteed by Definition \ref{def:broadcastside}  when $\frac{n}{m}\geq\tilde{\kappa}(1+2\delta)$ and $n$ is large enough. Receiver 1 then constructs $\hat{S}^m_{1}$ with
\begin{align*}
\hat{S}_{1}(t)=\psi(\tilde{S}_{2}(t),W(\hat{M}_1,t)),\quad t=1,\cdots,m,
\end{align*}
where $W(\hat{M}_1,t)$ is the $t$-th entry of $W^m_1(\hat{M}_1)$.
Receiver 2 sets $\hat{S}^m_{2}=\hat{S}^m_{2}(\hat{M}_2)$. It is easy to show that $(\tilde{S}^m_{1},\tilde{S}^m_{2},\hat{S}^m_{1})$ are jointly typical with respect to $p_{\tilde{S}_1,\tilde{S}_2,\hat{S}_1}$ with high probability, and $(\tilde{S}^m_{2},\hat{S}^m_{2})$ are jointly typical with respect to $p_{\tilde{S}_2,\hat{S}_2}$ with high probability. This completes the proof of the ``if" part.

Now we proceed to prove the ``only if" part. Consider an arbitrary tuple $(\tilde{\kappa},\tilde{\mathcal{Q}}_1,\tilde{\mathcal{Q}}_2)\in\tilde{\Gamma}$. Given any $\epsilon>0$,  according to Definition \ref{def:systemB}, we can find encoding function $f^{(m,n)}:\tilde{\mathcal{S}}^m_1\times\tilde{\mathcal{S}}^m_2\rightarrow\mathcal{X}^n$ as well as decoding functions $g_1^{(n,m)}:\mathcal{Y}^n_1\times\tilde{\mathcal{S}}^m_2\rightarrow\hat{\mathcal{S}}^m_1$ and $g_2^{(n,m)}:\mathcal{Y}^n_2\rightarrow\hat{\mathcal{S}}^m_2$ such that (\ref{eq:sysB1})-(\ref{eq:sysB3}) are satisfied.
Let $Q$ be  a random variable independent of $(\tilde{S}^m_{1},\tilde{S}^m_{2},X^n,Y^n_{1},Y^n_{2})$ and uniformly distributed over $\{1,\cdots,n\}$. Define $X=X(Q)$, $Y_i=Y_i(Q)$, $i=1,2$,  and $V=(V(Q),Q)$, where $V(t)=(Y^{t-1}_{1},Y^{n}_{2,t+1},\tilde{S}^m_{2})$ for all $t$. It is easy to verify that $V\leftrightarrow X\leftrightarrow(Y_1,Y_2)$ form a Markov chain.
Note that
\begin{align}
I(\tilde{S}^m_{1};\hat{S}^m_{1}|\tilde{S}^m_{2})&\leq I(\tilde{S}^m_{1};Y^n_{1}|\tilde{S}^m_{2})\nonumber\\
&\leq I(\tilde{S}^m_{1},\tilde{S}^m_{2};Y^n_{1})\nonumber\\
&\leq I(X^n;Y^n_{1})\nonumber\\
&=\sum\limits_{t=1}^n I(X^n;Y_{1}(t)|Y^{t-1}_{1})\nonumber\\
&\leq\sum\limits_{t=1}^n I(X^n,Y^{t-1}_{1};Y_1(t))\nonumber\\
&=\sum\limits_{t=1}^nI(X(t);Y_1(t))\nonumber\\
&=nI(X(Q);Y_1(Q)|Q)\nonumber\\
&\leq n(Q,X(Q);Y_1(Q))\nonumber\\
&=nI(X(Q);Y_1(Q))\nonumber\\
&=nI(X;Y_1)\label{eq:combc1}
\end{align}
and
\begin{align}
I(\tilde{S}^m_{2};\hat{S}^m_{2})&\leq I(\tilde{S}^m_{2};Y^n_{2})\nonumber\\
&=\sum\limits_{t=1}^nI(\tilde{S}^m_{2};Y_2(t)|Y^{n}_{2,t+1})\nonumber\\
&\leq\sum\limits_{t=1}^nI(Y^{t-1}_{1},Y^{n}_{2,t+1},\tilde{S}^m_{2};Y_2(t))\nonumber\\
&=\sum\limits_{t=1}^nI(V(t);Y_2(t))\nonumber\\
&=nI(V(Q);Y_2(Q)|Q)\nonumber\\
&\leq nI(V(Q),Q;Y_2(Q))\nonumber\\
&= nI(V;Y_2).\label{eq:combc2}
\end{align}
Moreover,
\begin{align}
&I(\tilde{S}^m_{1};\hat{S}^m_{1}|\tilde{S}^m_{2})+I(\tilde{S}^m_{2};\hat{S}^m_{2})\nonumber\\
&\leq I(\tilde{S}^m_{1};Y^n_{1}|\tilde{S}^m_{2})+I(\tilde{S}^m_{2};Y^n_{2})\nonumber\\
&=\sum\limits_{t=1}^n[I(\tilde{S}^m_{1};Y_1(t)|Y^{t-1}_{1},\tilde{S}^m_{2})+I(\tilde{S}^m_{2};Y_2(t)|Y^n_{2,t+1})]\nonumber\\
&\leq\sum\limits_{t=1}^n[I(X(t);Y_1(t)|Y^{t-1}_{1},\tilde{S}^m_{2})+I(\tilde{S}^m_{2};Y_2(t)|Y^n_{2,t+1})]\nonumber\\
&\leq\sum\limits_{t=1}^n[I(X(t),Y^n_{2,t+1};Y_1(t)|Y^{t-1}_{1},\tilde{S}^m_{2})\nonumber\\
&\hspace{0.4in}+I(Y^n_{2,t+1},\tilde{S}^m_{2};Y_2(t))]\nonumber\\
&=\sum\limits_{t=1}^n[I(X(t);Y_1(t)|Y^{t-1}_{1},Y^n_{2,t+1},\tilde{S}^m_{2})\nonumber\\
&\hspace{0.4in}+I(Y^n_{2,t+1};Y_1(t)|Y^{t-1}_{1},\tilde{S}^m_{2})+I(Y^n_{2,t+1},\tilde{S}^m_{2};Y_2(t))]\nonumber\\
&=\sum\limits_{t=1}^n[I(X(t);Y_1(t)|Y^{t-1}_{1},Y^n_{2,t+1},\tilde{S}^m_{2})\nonumber\\
&\hspace{0.4in}+I(Y^{t-1}_{1};Y_2(t)|Y^n_{2,t+1},\tilde{S}^m_{2})+I(Y^n_{2,t+1},\tilde{S}^m_{2};Y_2(t))]\label{eq:csiszar}\\
&=\sum\limits_{t=1}^n[I(X(t);Y_1(t)|Y^{t-1}_{1},Y^n_{2,t+1},\tilde{S}^m_{2})\nonumber\\
&\hspace{0.4in}+I(Y^{t-1}_{1},Y^n_{2,t+1},\tilde{S}^m_{2};Y_2(t))]\nonumber\\
&=\sum\limits_{t=1}^n[I(X(t);Y_1(t)|V(t))+I(V(t);Y_2(t))]\nonumber\\
&=n[I(X(Q);Y_1(Q)|V(Q),Q)+I(V(Q);Y_2(Q)|Q)]\nonumber\\
&\leq n[I(X(Q);Y_1(Q)|V(Q),Q)+I(V(Q),Q;Y_2(Q))]\nonumber\\
&= nI(X;Y_1|V)+nI(V;Y_2),\label{eq:combc3}
\end{align}
where (\ref{eq:csiszar}) follows by the Csisz\'{a}r sum identity \cite[p. 25]{EGK11}. Let $T$ be a random variable independent of $(\tilde{S}^m_{1}, \tilde{S}^m_{2},\hat{S}^m_{1},\hat{S}^m_{2})$ and uniformly distributed over $\{1,\cdots,m\}$. Define $\tilde{S}_i=\tilde{S}_i(T)$ and $\hat{S}^{(\epsilon)}_i=\hat{S}_i(T)$, $i=1,2$. Note that
\begin{align*}
p_{\tilde{S}_1,\tilde{S}_2,\hat{S}^{(\epsilon)}_1,\hat{S}^{(\epsilon)}_2}=\frac{1}{m}\sum\limits_{t=1}^mp_{\tilde{S}_1(t),\tilde{S}_2(t),\hat{S}_1(t),\hat{S}_2(t)}.
\end{align*}
Moreover, we have
\begin{align}
I(\tilde{S}^m_{1};\hat{S}^m_{1}|\tilde{S}^m_{2})&=\sum\limits_{t=1}^mI(\tilde{S}_1(t);\hat{S}^m_{1}|\tilde{S}^{t-1}_{1},\tilde{S}^m_{2})\nonumber\\
&=\sum\limits_{t=1}^mI(\tilde{S}_1(t);\hat{S}^m_{1},\tilde{S}^{t-1}_{1},\tilde{S}^{t-1}_{2},\tilde{S}^n_{2,t+1}|\tilde{S}_2(t))\nonumber\\
&\geq\sum\limits_{t=1}^mI(\tilde{S}_1(t);\hat{S}_1(t)|\tilde{S}_2(t))\nonumber\\
&=mI(\tilde{S}_1(T);\hat{S}_1(T)|\tilde{S}_2(T),T)\nonumber\\
&=mI(\tilde{S}_1(T);\hat{S}_1(T),T|\tilde{S}_2(T))\nonumber\\
&\geq mI(\tilde{S}_1(T);\hat{S}_1(T)|\tilde{S}_2(T))\nonumber\\
&= mI(\tilde{S}_1;\hat{S}^{(\epsilon)}_1|\tilde{S}_2)\label{eq:combs1}
\end{align}
and
\begin{align}
I(\tilde{S}^m_{2};\hat{S}^m_{2})&=\sum\limits_{t=1}^mI(\tilde{S}_2(t);\hat{S}^m_{2}|\tilde{S}^{t-1}_{2})\nonumber\\
&=\sum\limits_{t=1}^mI(\tilde{S}_2(t);\hat{S}^m_{2},\tilde{S}^{t-1}_{2})\nonumber\\
&\geq\sum\limits_{t=1}^mI(\tilde{S}_2(t);\hat{S}_2(t))\nonumber\\
&=mI(\tilde{S}_2(T);\hat{S}_2(T)|T)\nonumber\\
&=mI(\tilde{S}_2(T);\hat{S}_2(T),T)\nonumber\\
&\geq mI(\tilde{S}_2(T);\hat{S}_2(T))\nonumber\\
&=mI(\tilde{S}_2;\hat{S}^{(\epsilon)}_2).\label{eq:combs2}
\end{align}
It follows by (\ref{eq:combc1}), (\ref{eq:combc2}), (\ref{eq:combc3}), (\ref{eq:combs1}), and (\ref{eq:combs2}) that
\begin{align*}
(I(\tilde{S}_1;\hat{S}^{(\epsilon)}_1|\tilde{S}_2),I(\tilde{S}_2;\hat{S}^{(\epsilon)}_2))\in\frac{n}{m}\mathcal{C}_{1}(p_{Y_1,Y_2|X}).
\end{align*}
Since $\{p_{\tilde{S}_1,\tilde{S}_2,\hat{S}^{(\epsilon)}_1,\hat{S}^{(\epsilon)}_2}:\epsilon>0\}$ is contained in a compact set and
\begin{align*}
&\min_{\tilde{q}_1\in\tilde{\mathcal{Q}}_1}\|p_{\tilde{S}_1,\tilde{S}_2,\hat{S}^{(\epsilon)}_1}-\tilde{q}_1\|\leq\epsilon,\\
&\min_{\tilde{q}_2\in\tilde{\mathcal{Q}}_2}\|p_{\tilde{S}_2,\hat{S}^{(\epsilon)}_2}-\tilde{q}_2\|\leq\epsilon
 \end{align*}
 for every $\epsilon>0$, $i=1,2$, one can find a sequence $\epsilon_1, \epsilon_2, \cdots$ converging to zero such that
\begin{align*}
\lim\limits_{k\rightarrow\infty}p_{\tilde{S}_1,\tilde{S}_2,\hat{S}^{(\epsilon_k)}_1,\hat{S}^{(\epsilon_k)}_2}= p_{\tilde{S}_1,\tilde{S}_2,\hat{S}_1,\hat{S}_2}
\end{align*}
for some $p_{\tilde{S}_1,\tilde{S}_2,\hat{S}_1,\hat{S}_2}$ with $p_{\tilde{S}_1,\tilde{S}_2,\hat{S}_1}\in\tilde{\mathcal{Q}}_1$ and $p_{\tilde{S}_2,\hat{S}_2}\in\tilde{\mathcal{Q}}_2$. It is clear that
\begin{align*}
&I(\tilde{S}_1;\hat{S}_1|\tilde{S}_2)\geq R_{\tilde{S}_1|\tilde{S}_2}(\tilde{\mathcal{Q}}_1),\\
&I(\tilde{S}_2;\hat{S}_2)\geq R_{\tilde{S}_2}(\tilde{\mathcal{Q}}_2).
\end{align*}
Now the proof can be completed via a simple limiting argument.
\end{IEEEproof}

\section{A Necessary Condition for the Source Broadcast Problem}\label{sec:application}

\subsection{Necessary Condition}

We shall show that the source-channel separation theorem for System $\tilde{\Pi}$ (i.e.,  Theorem \ref{thm:systemB}) can be leveraged to establish a necessary condition for System $\Pi$ via a simple reduction argument. Let $\mathcal{R}_1(p_{S,\hat{S}_1,\hat{S}_2})$ denote the set of $(R_1,R_2)\in\mathbb{R}^2_+$ satisfying
\begin{align*}
&R_1\leq I(S;\hat{S}_1|U),\\
&R_2\leq I(U;\hat{S}_2)
\end{align*}
for some $p_{U,S,\hat{S}_1,\hat{S}_2}=p_{U|S}p_{S,\hat{S}_1,\hat{S}_2}$. Similarly, let $\mathcal{R}_2(p_{S,\hat{S}_1,\hat{S}_2})$ denote the set of $(R_1,R_2)\in\mathbb{R}^2_+$ satisfying
\begin{align*}
&R_1\leq I(U;\hat{S}_1),\\
&R_2\leq I(S;\hat{S}_2|U)
\end{align*}
for some $p_{U,S,\hat{S}_1,\hat{S}_2}=p_{U|S}p_{S,\hat{S}_1,\hat{S}_2}$.

\begin{theorem}\label{thm:remote}
For any $(\kappa,\mathcal{Q}_1,\mathcal{Q}_2)\in\Gamma$, there exists $p_{S,\hat{S}_1,\hat{S}_2}$ with $p_{S,\hat{S}_i}\in\mathcal{Q}_i$, $i=1,2$, such that
\begin{align}
\mathcal{R}_i(p_{S,\hat{S}_1,\hat{S}_2})\subseteq\kappa\mathcal{C}_i(p_{Y_1,Y_2|X}),\quad i=1,2.\label{eq:tp1}
\end{align}
\end{theorem}
\begin{IEEEproof}
By symmetry, it suffices to prove (\ref{eq:tp1}) for $i=1$. We augment the probability space by introducing a remote source  $\{(\tilde{S}_{1}(t),\tilde{S}_{2}(t))\}_{t=1}^\infty$ such that $(\tilde{S}_{1}(t),\tilde{S}_{2}(t),S(t))$, $t=1,2,\cdots$, are independent and identically distributed over finite alphabet $\tilde{\mathcal{S}}_1\times\tilde{\mathcal{S}}_2\times\mathcal{S}$. Consider an arbitrary tuple $(\kappa,\mathcal{Q}_1,\mathcal{Q}_2)\in\Gamma$. Given any $\epsilon>0$, according to Definition \ref{def:systemA}, we can find encoding function $f^{(m,n)}:\mathcal{S}^m\rightarrow\mathcal{X}^n$ and decoding functions $g_i^{(n,m)}:\mathcal{Y}^n_i\rightarrow\hat{\mathcal{S}}^m_i$, $i=1,2$, satisfying (\ref{eq:maincond1}) and (\ref{eq:maincond2}). Let $T$ be a random variable independent of $(\tilde{S}^m_{1},\tilde{S}^m_{2},S^m,\hat{S}^m_{1},\hat{S}^m_{2})$ and uniformly distributed over $\{1,\cdots,m\}$. Define $\tilde{S}_i=\tilde{S}_i(T)$, $i=1,2$, $S=S(T)$, and $\hat{S}^{(\epsilon)}_i=\hat{S}_i(T)$, $i=1,2$. It is clear that the distribution of $(\tilde{S}_1,\tilde{S}_2,S)$ is identical with that of $(\tilde{S}_1(t),\tilde{S}_2(t),S(t))$ for every $t$, and $(\tilde{S}_1,\tilde{S}_2)\leftrightarrow S\leftrightarrow (\hat{S}^{(\epsilon)}_1,\hat{S}^{(\epsilon)}_2)$ form a Markov chain. Moreover, we have
\begin{align*}
\frac{1}{m}\sum\limits_{t=1}^mp_{\tilde{S}_1(t),\tilde{S}_2(t),S(t),\hat{S}_1(t),\hat{S}_2(t)}=p_{\tilde{S}_1,\tilde{S}_2,S,\hat{S}^{(\epsilon)}_1,\hat{S}^{(\epsilon)}_2}.
\end{align*}
Since $\min_{q_i\in\mathcal{Q}_i}\|p_{S,\hat{S}^{(\epsilon)}_i}-q_i\|\leq \epsilon$ for every $\epsilon>0$, $i=1,2$, one can find a sequence $\epsilon_1, \epsilon_2, \cdots$ converging to zero such that
\begin{align}
\lim\limits_{k\rightarrow\infty}p_{\tilde{S}_1,\tilde{S}_2,S,\hat{S}^{(\epsilon_k)}_1,\hat{S}^{(\epsilon_k)}_2}= p_{\tilde{S}_1,\tilde{S}_2,S,\hat{S}_1,\hat{S}_2}\label{eq:reduction}
\end{align}
for some $p_{\tilde{S}_1,\tilde{S}_2,S,\hat{S}_1,\hat{S}_2}$ with $p_{S,\hat{S}_i}\in\mathcal{Q}_i$, $i=1,2$. Note that (\ref{eq:reduction}) implies $(\kappa,\{p_{\tilde{S}_1,\tilde{S}_2,\hat{S}_1}\},\{p_{\tilde{S}_2,\hat{S}_2}\})\in\tilde{\Gamma}$. Therefore, it follows from Theorem \ref{thm:systemB} that
\begin{align*}
(I(\tilde{S}_1;\hat{S}_1|\tilde{S}_2),I(\tilde{S}_2;\hat{S}_2))\in\kappa\mathcal{C}_{1}(p_{Y_1,Y_2|X}).
\end{align*}
Here one can fix $p_{S,\hat{S}_1,\hat{S}_2}$ and choose $p_{\tilde{S}_1,\tilde{S}_2|S}$ arbitrarily. Since $I(\tilde{S}_1;\hat{S}_1|\tilde{S}_2)\leq I(S;\hat{S}_1|\tilde{S}_2)$, there is no loss of generality in setting $\tilde{S}_1=S$. Denoting $\tilde{S}_2$ by $U$ completes the proof of Theorem \ref{thm:remote}.
\end{IEEEproof}

Remark: Since  $\mathcal{C}_{1}(p_{Y_1,Y_2|X})$ and $\mathcal{C}_{2}(p_{Y_1,Y_2|X})$ are convex sets, it follows that (\ref{eq:tp1}) holds if and only if $\kappa\mathcal{C}_i(p_{Y_1,Y_2|X})$ contains all the extreme points of $\mathcal{R}_i(p_{S,\hat{S}_1,\hat{S}_2})$, $i=1,2$. One can show via a standard application of the support lemma\cite[p. 631]{EGK11} that, in contrast with the cardinality bound $|\mathcal{U}|\leq|\mathcal{S}|+1$ for preserving $\mathcal{R}_i(p_{S,\hat{S}_1,\hat{S}_2})$, $i=1,2$, it suffices to have $|\mathcal{U}|\leq|\mathcal{S}|$ for the purpose of realizing all their extreme points.



\subsection{The Binary Uniform Source with the Hamming Distortion Measure}

In this subsection we set $\mathcal{S}=\hat{\mathcal{S}}_1=\hat{\mathcal{S}}_2=\{0,1\}$, $p_S(0)=p_S(1)=\frac{1}{2}$, and $w_1=w_2=w_H$, where $w_H$ is the Hamming distortion measure, i.e.,
\begin{align*}
w_H(s,\hat{s})=\left\{
                 \begin{array}{ll}
                   0, & s=\hat{s} \\
                   1, & \mbox{otherwise}
                 \end{array}
               \right..
\end{align*}
The problem is trivial\footnote{In fact, it reduces to a point-to-point problem.}  when $d_1=\frac{1}{2}$ or $d_2=\frac{1}{2}$.  Therefore, we shall focus on the non-degenerate case $d_i\in[0,\frac{1}{2})$, $i=1,2$, and assume
\begin{align*}
C(p_{Y_i|X})\triangleq\max\limits_{p_X}I(X;Y_i)>0,\quad i=1,2,
\end{align*}
correspondingly.



\begin{proposition}\label{lem:R12}
If $p_{S,\hat{S}_1,\hat{S}_2}$ is such that $\mathbb{E}[w_H(S,\hat{S}_i)]\leq d_i$, $i=1,2$, with $d_1\leq d_2$, then
\begin{align}
&\mathcal{R}_1(p_{S,\hat{S}_1,\hat{S}_2})\supseteq\mathcal{C}(\mbox{BS-BC}(d_1,d_2)),\label{eq:contain1}\\
&\mathcal{R}_2(p_{S,\hat{S}_1,\hat{S}_2})\supseteq\tilde{\mathcal{C}}(\mbox{BS-BC}(d_1,d_2)),\label{eq:contain2}
\end{align}
where $\mathcal{C}(\mbox{BS-BC}(d_1,d_2))$ (see Section \ref{sec:IIIexample} for its definition) is given by the set of $(R_1,R_2)\in\mathbb{R}^+_2$  satisfying
\begin{align*}
&R_1\leq H_b(\alpha*d_1)-H_b(d_1),\\
&R_2\leq 1-H_b(\alpha*d_2)
\end{align*}
for some $\alpha\in[0,\frac{1}{2}]$, and
$\tilde{\mathcal{C}}(\mbox{BS-BC}(d_1,d_2))$ is given by the set of $(R_1,R_2)\in\mathbb{R}^+_2$ satisfying
\begin{align*}
&R_1\leq\beta[1-H_b(d_1)],\\
&R_2\leq(1-\beta)[1-H_b(d_2)]
\end{align*}
for some $\beta\in[0,1]$. Moreover,
\begin{align}
&\mathcal{R}_1(p_{S,\hat{S}_1,\hat{S}_2})=\mathcal{C}(\mbox{BS-BC}(d_1,d_2)),\label{eq:equal1}\\
&\mathcal{R}_2(p_{S,\hat{S}_1,\hat{S}_2})=\tilde{\mathcal{C}}(\mbox{BS-BC}(d_1,d_2))\label{eq:equal2}
\end{align}
when $p_{\hat{S}_1,\hat{S}_2|S}$ is a $\mbox{BS-BC}(d_1,d_2)$ with $d_1\leq d_2$.
\end{proposition}
\begin{IEEEproof}
Let $p_{U,S,\hat{S}_1,\hat{S}_2}=p_{U|S}p_{S,\hat{S}_1,\hat{S}_2}$, where $p_{U|S}$ is a $\mbox{BSC}(\alpha)$ with $\alpha\in[0,\frac{1}{2}]$.
We have
\begin{align}
&\min\limits_{p_{\hat{S}_1|S}:\mathbb{E}[w_H(S,\hat{S}_1)]\leq d_1}I(S;\hat{S}_1|U)\nonumber\\
&=\min\limits_{p_{\hat{S}_1|S}:\mathbb{E}[w_H(S,\hat{S}_1)]\leq d_1}I(S;\hat{S}_1)-I(U;\hat{S}_1)\nonumber\\
&=\min\limits_{p_{\hat{S}_1|S}:\mathbb{E}[w_H(S,\hat{S}_1)]\leq d_1}H(U|\hat{S}_1)-H(S|\hat{S}_1)\label{eq:sameentropy}\\
&=\min\limits_{d'_1\in[0,d_1]}H_b(\alpha*d'_1)-H_b(d'_1)\label{eq:cyclic}\\
&=H_b(\alpha*d_1)-H_b(d_1),\label{eq:monotonicity}
\end{align}
where (\ref{eq:sameentropy}) follows since $H(S)=H(U)=1$, (\ref{eq:cyclic}) follows from \cite[Lemma 2]{SSC13}, and (\ref{eq:monotonicity}) is due to the fact that $H_b(\alpha*d'_1)-H_b(d'_1)$ is a monotonically decreasing function of $d'_1$ for $d'_1\in[0,\frac{1}{2}]$. Similarly, it can be shown that
\begin{align}
\min\limits_{p_{\hat{S}_2|S}:\mathbb{E}[w_H(S,\hat{S}_2)]\leq d_2}I(U;\hat{S}_2)=1-H_b(\alpha*d_2).\label{eq:monotonicity2}
\end{align}
Combining (\ref{eq:monotonicity}) and (\ref{eq:monotonicity2}) proves (\ref{eq:contain1}).

It is easy to see that $(I(S;\hat{S}_1),0)$ and $(0,I(S;\hat{S}_2))$ are contained in $\mathcal{R}_2(p_{S,\hat{S}_1,\hat{S}_2})$. Note that
\begin{align*}
I(S;\hat{S}_i)\geq 1-H_b(d_i)
\end{align*}
if $\mathbb{E}[w_H(S,\hat{S}_i)]\leq d_i$, $i=1,2$. Now one can readily prove (\ref{eq:contain2}) by invoking the fact that $\mathcal{R}_2(p_{S,\hat{S}_1,\hat{S}_2})$ is a convex set.

Since (\ref{eq:equal1}) is obviously true, only (\ref{eq:equal2}) remains to be proved. If  $p_{\hat{S}_1,\hat{S}_2|S}$ is a $\mbox{BS-BC}(d_1,d_2)$ with $d_1\leq d_2$, then, for any $\lambda\in[0,1]$,
\begin{align*}
&\lambda I(U;\hat{S}_1)+(1-\lambda)I(S;\hat{S}_2|U)\\
&=\lambda (1-H(\hat{S}_1|U))+(1-\lambda)[H(\hat{S}_2|U)-H_b(d_2)]\\
&\leq\max\limits_{u\in\mathcal{U}}\lambda (1-H(\hat{S}_1|U=u))\\
&\hspace{0.5in}+(1-\lambda)[H(\hat{S}_2|U=u)-H_b(d_2)]\\
&\leq\max\limits_{\alpha\in[0,\frac{1}{2}]}\lambda (1-H_b(\alpha*d_1))\\
&\hspace{0.6in}+(1-\lambda)[H_b(\alpha*d_2)-H_b(d_2)].
\end{align*}
Define $v=H_b(\alpha*d_1)$, which is a monotonically increasing function of $\alpha$. Note that
\begin{align*}
&\lambda (1-H_b(\alpha*d_1))+(1-\lambda)[H_b(\alpha*d_2)-H_b(d_2)]\\
&=\lambda (1-v)+(1-\lambda)[H_b(H^{-1}_b(v)*d)-H_b(d_2)],
\end{align*}
where $d=\frac{d_2-d_1}{1-2d_1}$. It follows by the convexity of $H_b(H^{-1}_b(v)*d)$ in $v$ \cite[Lemma 2]{WZ73} that
\begin{align}
&\max\limits_{\alpha\in[0,\frac{1}{2}]}\lambda (1-H_b(\alpha*d_1))+(1-\lambda)[H_b(\alpha*d_2)-H_b(d_2)]\nonumber\\
&=\max\limits_{\alpha\in\{0,\frac{1}{2}\}}\lambda (1-H_b(\alpha*d_1))\nonumber\\
&\hspace{0.6in}+(1-\lambda)[H_b(\alpha*d_2)-H_b(d_2)].\label{eq:uselater}
\end{align}
Therefore, we must have $\mathcal{R}_2(p_{S,\hat{S}_1,\hat{S}_2})\subseteq\tilde{\mathcal{C}}(\mbox{BS-BC}(d_1,d_2))$, which together with (\ref{eq:contain2}), proves (\ref{eq:equal2}).
\end{IEEEproof}

Remark: The proof of Proposition \ref{lem:R12} indicates that, for the binary uniform source with the Hamming distortion measure, there is no loss of optimality (as far as Theorem \ref{thm:remote} is concerned) in restricting $p_{U|S}$ to be a binary symmetric channel, which provides a certain justification for the choice of the auxiliary random variable in \cite{TKS13}.

Note that the rate pairs $(C(p_{Y_1|X}),0)$ and $(0,C(p_{Y_2|X}))$ are contained in both $\mathcal{C}_{1}(p_{Y_1,Y_2|X})$ and $\mathcal{C}_{2}(p_{Y_1,Y_2|X})$.
It is easy to see that $\mathcal{C}(\mbox{BS-BC}(d_1,d_2))\subseteq\kappa\mathcal{C}_{1}(p_{Y_1,Y_2|X})$ implies
\begin{align*}
1-H_b(d_i)\leq\kappa C(p_{Y_i|X}),\quad i=1,2,
\end{align*}
which further implies $\tilde{\mathcal{C}}(\mbox{BS-BC}(d_1,d_2))\subseteq\kappa\mathcal{C}_{2}(p_{Y_1,Y_2|X})$ when $d_1\leq d_2$. This observation, together with Proposition \ref{lem:R12}, shows that, for the binary uniform source with the Hamming distortion measure, Theorem \ref{thm:remote} is equivalent to the following more explicit result.

\begin{theorem}\label{thm:binary}
 For any $(\kappa,\mathcal{Q}(w_H,d_1),\mathcal{Q}(w_H,d_2))\in\Gamma$ with $d_1\leq d_2$,
  \begin{align*}
  \mathcal{C}(\mbox{BS-BC}(d_1,d_2))\subseteq\kappa\mathcal{C}_{1}(p_{Y_1,Y_2|X}).
  \end{align*}
    By symmetry,  for any $(\kappa,\mathcal{Q}(w_H,d_1),\mathcal{Q}(w_H,d_2))\in\Gamma$ with $d_1\geq d_2$,
    \begin{align*}
    \mathcal{C}(\mbox{BS-BC}(d_1,d_2))\subseteq\kappa\mathcal{C}_{2}(p_{Y_1,Y_2|X}).
    \end{align*}
\end{theorem}

Define $\kappa^{\star}=\min\{\kappa\geq 0:\mathcal{C}(\mbox{BS-BC}(d_1,d_2))\subseteq\kappa\mathcal{C}_{1}(p_{Y_1,Y_2|X})\}$ if $d_1\leq d_2$, and $\kappa^{\star}=\min\{\kappa\geq 0:\mathcal{C}(\mbox{BS-BC}(d_1,d_2))\subseteq\kappa\mathcal{C}_{2}(p_{Y_1,Y_2|X})\}$ if $d_1\geq d_2$. It is obvious that
\begin{align}
\kappa^{\star}\geq\kappa^{\dagger}\triangleq\max\Big\{\frac{1-H_b(d_1)}{C(p_{Y_1|X})},\frac{1-H_b(d_2)}{C(p_{Y_2|X})}\Big\},\label{eq:obvious}
\end{align}
i.e., the necessary condition stated in Theorem \ref{thm:binary} is at least as strong as the one implied by the source-channel separation theorem for point-to-point communication systems.
We shall show that in some cases it is possible to determine whether $\kappa^{\star}$ is equal to or strictly greater than $\kappa^{\dagger}$ without an explicit characterization of $\mathcal{C}_{i}(p_{Y_1,Y_2|X})$, $i=1,2$.

Recall that  $\mathcal{C}(\mbox{BS-BC}(d_1,d_2))$ with $d_1\leq d_2$ is given by the the set of $(R_1,R_2)\in\mathbb{R}^2_+$ satisfying
\begin{align}
&R_1\leq R_1(\alpha)\triangleq H_b(\alpha*d_1)-H_b(d_1),\label{eq:bsbcexp1}\\
&R_2\leq R_2(\alpha)\triangleq 1-H_b(\alpha*d_2)\label{eq:bsbcexp2}
\end{align}
for some $\alpha\in[0,\frac{1}{2}]$. 
It can be verified that\footnote{We set $\frac{(1-2d_2)\log\Big(\frac{1-d_2}{d_2}\Big)}{(1-2d_1)\log\Big(\frac{1-d_1}{d_1}\Big)}=1$ when $d_1=d_2=0$.}
\begin{align}
&\left.\frac{\mathrm{d}R_2(\alpha)}{\mathrm{d}R_1(\alpha)}\right|_{\alpha=0}=-\frac{(1-2d_2)\log\Big(\frac{1-d_2}{d_2}\Big)}{(1-2d_1)\log\Big(\frac{1-d_1}{d_1}\Big)},\label{eq:derivative0}\\
&\left.\frac{\mathrm{d}R_2(\alpha)}{\mathrm{d}R_1(\alpha)}\right|_{\alpha=\frac{1}{2}}=-\frac{(1-2d_2)^2}{(1-2d_1)^2}.\label{eq:derivativehalf}
\end{align}
In view of the fact that $\frac{\mathrm{d}R_2(\alpha)}{\mathrm{d}R_1(\alpha)}$ is a monotonically decreasing function of $\alpha$ for $\alpha\in[0,\frac{1}{2}]$, it is clear that
\begin{align*}
&\mathcal{C}(\mbox{BS-BC}(d_1,d_2))\\
&\subseteq\kappa\Big\{(R_1,R_2)\in\mathbb{R}^2_+:\frac{R_1}{C(p_{Y_1|X})}+\frac{R_2}{C(p_{Y_2|X})}\leq 1\Big\}
\end{align*}
if one of the following conditions are satisfied:
\begin{enumerate}
\item $1-H_b(d_1)\leq\kappa C(p_{Y_1|X})$ and $\frac{(1-2d_1)^2}{(1-2d_2)^2}\geq\frac{C(p_{Y_1|X})}{C(p_{Y_2|X})}$,

\item $1-H_b(d_2)\leq\kappa C(p_{Y_2|X})$ and $\frac{(1-2d_1)\log\Big(\frac{1-d_1}{d_1}\Big)}{(1-2d_2)\log\Big(\frac{1-d_2}{d_2}\Big)}\leq\frac{C(p_{Y_1|X})}{C(p_{Y_2|X})}$.
\end{enumerate}
This observation, together with (\ref{eq:obvious}) as well as the fact that
\begin{align*}
&\Big\{(R_1,R_2)\in\mathbb{R}^2_+:\frac{R_1}{C(p_{Y_1|X})}+\frac{R_2}{C(p_{Y_2|X})}\leq 1\Big\}\\
&\subseteq\mathcal{C}_1(p_{Y_1,Y_2|X}),
\end{align*}
yields the following result.
\begin{proposition}\label{prop:trivial}
If $d_1\leq d_2$, then
\begin{align*}
\kappa^{\star}=\kappa^{\dagger}
=\left\{
   \begin{array}{ll}
     \frac{1-H_b(d_1)}{C(p_{Y_1|X})}, & \frac{(1-2d_1)^2}{(1-2d_2)^2}\geq\frac{C(p_{Y_1|X})}{C(p_{Y_2|X})} \\
     \frac{1-H_b(d_2)}{C(p_{Y_2|X})}, & \frac{(1-2d_1)\log\Big(\frac{1-d_1}{d_1}\Big)}{(1-2d_2)\log\Big(\frac{1-d_2}{d_2}\Big)}\leq\frac{C(p_{Y_1|X})}{C(p_{Y_2|X})}
   \end{array}
 \right..
\end{align*}
By symmetry, if $d_1\geq d_2$, then
\begin{align*}
\kappa^{\star}=\kappa^{\dagger}
=\left\{
   \begin{array}{ll}
     \frac{1-H_b(d_2)}{C(p_{Y_2|X})}, & \frac{(1-2d_2)^2}{(1-2d_1)^2}\geq\frac{C(p_{Y_2|X})}{C(p_{Y_1|X})} \\
     \frac{1-H_b(d_1)}{C(p_{Y_1|X})}, & \frac{(1-2d_2)\log\Big(\frac{1-d_2}{d_2}\Big)}{(1-2d_1)\log\Big(\frac{1-d_1}{d_1}\Big)}\leq\frac{C(p_{Y_2|X})}{C(p_{Y_1|X})}
   \end{array}
 \right..
\end{align*}
\end{proposition}

Remark: A simple sufficient condition for $(\kappa,\mathcal{Q}(w_H,d_1),\mathcal{Q}(w_H,d_2))\in\Gamma$ is that
\begin{align*}
\max\{1-H_b(d_1),1-H_b(d_2)\}\leq \kappa C(p_{Y_1|X},p_{Y_2|X}),
\end{align*}
where $C(p_{Y_1|X},p_{Y_2|X})\triangleq\max_{p_X}\min\{I(X;Y_1),I(X;Y_2)\}$ is the capacity of the compound channel $\{p_{Y_1|X},p_{Y_2|X}\}$. Proposition \ref{prop:trivial} indicates that this sufficient condition is also necessary when $C(p_{Y_1|X},p_{Y_2|X})=C(p_{Y_1|X})$ and $d_1\leq d_2$ (or  $C(p_{Y_1|X},p_{Y_2|X})=C(p_{Y_2|X})$ and $d_1\geq d_2$). For the special case $d_1=d_2=d$, it can be shown that $(\kappa,\mathcal{Q}(w_H,d),\mathcal{Q}(w_H,d))\in\Gamma$ if and only if
\begin{align*}
1-H_b(d)\leq \kappa C(p_{Y_1|X},p_{Y_2|X}).
\end{align*}
On the other hand, for this special case, Proposition \ref{prop:trivial} gives
\begin{align*}
\kappa^{\star}=\kappa^{\dagger}=\max\Big\{\frac{1-H_b(d)}{C(p_{Y_1|X})},\frac{1-H_b(d)}{C(p_{Y_2|X})}\Big\}.
\end{align*}
Since $C(p_{Y_1|X},p_{Y_2|X})$ can be strictly smaller than $\min\{C(p_{Y_1|X}),C(p_{Y_2|X})\}$, the necessary condition stated in Theorem \ref{thm:binary} is not sufficient in general.





For every $R_1\in[0,C(p_{Y_1|X})]$, we set
\begin{align*}
\phi(R_1)=\max\{R_2:(R_1,R_2)\in\mathcal{C}_1(p_{Y_1,Y_2|X})\}.
\end{align*}
Note that $\phi:[0,C(p_{Y_1|X})]\rightarrow[0,C(p_{Y_2|X})]$ is monotonically decreasing and concave. Define
\begin{align*}
&\phi'_+(0)=\lim\limits_{R_1\downarrow 0}\frac{C(p_{Y_2|X})-\phi(R_1)}{R_1},\\
&\phi'_-(C(p_{Y_1|X}))=\lim\limits_{R_1\uparrow C(p_{Y_1|X})}\frac{\phi(R_1)}{C(p_{Y_1|X})-R_1}.
\end{align*}
Similarly, we  set
\begin{align*}
\varphi(R_2)=\max\{R_1:(R_1,R_2)\in\mathcal{C}_2(p_{Y_1,Y_2|X})\}
\end{align*}
for every $R_2\in[0,C(p_{Y_2|X})]$, and define
\begin{align*}
&\varphi'_+(0)=\lim\limits_{R_2\downarrow 0}\frac{C(p_{Y_1|X})-\varphi(R_2)}{R_2},\\
&\varphi'_-(C(p_{Y_2|X}))=\lim\limits_{R_2\uparrow C(p_{Y_2|X})}\frac{\varphi(R_2)}{C(p_{Y_2|X})-R_2}.
\end{align*}
Now consider the case $d_1\leq d_2$. It is clear that we must have $1-H_b(d_1)<\kappa^{\star}C(p_{Y_1|X})$ if
\begin{align}
\frac{(1-2d_2)^2}{(1-2d_1)^2}>\phi'_-(C(p_{Y_1|X}));\label{eq:sat1}
\end{align}
similarly, we must have $1-H_b(d_2)<\kappa^{\star}C(p_{Y_2|X})$ if
\begin{align}
\frac{(1-2d_2)\log\Big(\frac{1-d_2}{d_2}\Big)}{(1-2d_1)\log\Big(\frac{1-d_1}{d_1}\Big)}<\phi'_+(0);\label{eq:sat2}
\end{align}
moreover, since $\phi'_+(0)\leq\phi'_-(C(p_{Y_1|X}))$, it follows that (\ref{eq:sat1}) and (\ref{eq:sat2}) cannot be satisfied simultaneously when $d_1=d_2$.
The following result is a simple consequence of this observation.
\begin{proposition}\label{prop:nontrivial}
When $d_1<d_2$, we have $\kappa^{\star}>\kappa^{\dagger}$ if
\begin{align*}
&\frac{(1-2d_2)\log\Big(\frac{1-d_2}{d_2}\Big)}{(1-2d_1)\log\Big(\frac{1-d_1}{d_1}\Big)}<\phi'_+(0),\\
&\frac{(1-2d_2)^2}{(1-2d_1)^2}>\phi'_-(C(p_{Y_1|X})).
\end{align*}
By symmetry, when $d_1> d_2$, we have $\kappa^{\star}>\kappa^{\dagger}$ if
\begin{align*}
&\frac{(1-2d_1)\log\Big(\frac{1-d_1}{d_1}\Big)}{(1-2d_2)\log\Big(\frac{1-d_2}{d_2}\Big)}<\varphi'_+(0),\\
&\frac{(1-2d_1)^2}{(1-2d_2)^2}>\varphi'_-(C(p_{Y_2|X})).
\end{align*}
\end{proposition}

A channel $p_{Y|X}:\mathcal{X}\rightarrow\mathcal{Y}$ with $\mathcal{X}=\{0,1,\cdots,M-1\}$ for some integer $M\geq 2$ is said to be circularly symmetric \cite[Definition 1]{WKP07} (see also \cite[Definition 4]{Nair10}) if there exists a bijective function $\mu:\mathcal{Y}\rightarrow\mathcal{Y}$ such that $\mu^M(y)=y$ and $p_{Y|X}(\mu^x(y)|x)=p_{Y|X}(y|0)$ for all $(x,y)\in\mathcal{X}\times\mathcal{Y}$, where $\mu^k$ denotes the $k$-times self-composition of $\mu$ (with $\mu^0$ being the identity function). Note that the binary symmetric channel is circularly symmetric with $\mu:\{0,1\}\rightarrow\{0,1\}$ given by
\begin{align*}
\mu(y)=\left\{
      \begin{array}{ll}
        1, & y=0 \\
        0, & y=1
      \end{array}
    \right.;
\end{align*}
the binary erasure channel is also circularly symmetric, and the associated $\mu:\{0,1,e\}\rightarrow\{0,1,e\}$ is given by
\begin{align*}
\mu(y)=\left\{
         \begin{array}{ll}
           1, & y=0 \\
           0, & y=1 \\
           e, & y=e
         \end{array}
       \right..
\end{align*}


\begin{proposition}\label{prop:symmetry}
If both $p_{Y_1|X}$ and $p_{Y_2|X}$ are circularly symmetric, then
\begin{align*}
  \kappa^{\star}=\min\{\kappa\geq 0:\mathcal{C}(\mbox{BS-BC}(d_1,d_2))\subseteq\kappa\mathcal{C}(p_{Y_1,Y_2|X})\}.
  \end{align*}
\end{proposition}
\begin{IEEEproof}
By symmetry, it suffices to consider the case $d_1\leq d_2$. Let $\mathcal{C}_{\mbox{sc}}(p_{Y_1,Y_2|X})$ denote the superposition coding inner bound of $\mathcal{C}(p_{Y_1,Y_2|X})$, i.e., the set of $(R_1,R_2)\in\mathbb{R}^2_+$ satisfying
\begin{align*}
&R_2\leq I(V;Y_2),\\
&R_1+R_2\leq I(X;Y_1|V)+I(V;Y_2),\\
&R_1+R_2\leq I(X;Y_1)
\end{align*}
for some $p_{V,X,Y_1,Y_2}=p_{V,X}p_{Y_1,Y_2|X}$. In light of \cite[Lemma 2]{Nair10}, the uniform distribution on $\mathcal{X}$ forms a sufficient class of distributions for broadcast channel $p_{Y_1,Y_2|X}$ if both $p_{Y_1|X}$ and $p_{Y_2|X}$ are circularly symmetric. As a consequence, one can readily show that
\begin{align*}
&\mathcal{C}_{\mbox{sc}}(p_{Y_1,Y_2|X})\\
&=\mathcal{C}_1(p_{Y_1,Y_2|X})\cap\{(R_1,R_2):R_1+R_2\leq C(p_{Y_1|X})\}.
\end{align*}
Note that, if $\mathcal{C}(\mbox{BS-BC}(d_1,d_2))\subseteq\kappa\mathcal{C}_1(p_{Y_1,Y_2|X})$, then we must have
\begin{align*}
1-H_b(d_1)\leq\kappa C(p_{Y_1|X}),
\end{align*}
which, together with the fact that $\frac{\mathrm{d}R_2(\alpha)}{\mathrm{d}R_1(\alpha)}\in[-1,0]$ for $\alpha\in[0,\frac{1}{2}]$, implies
\begin{align*}
\mathcal{C}(\mbox{BS-BC}(d_1,d_2))\subseteq\kappa\{(R_1,R_2):R_1+R_2\leq C(p_{Y_1|X})\}.
\end{align*}
Therefore,
\begin{align*}
&\mathcal{C}(\mbox{BS-BC}(d_1,d_2))\subseteq\kappa\mathcal{C}_1(p_{Y_1,Y_2|X})\\
&\Rightarrow\mathcal{C}(\mbox{BS-BC}(d_1,d_2))\subseteq\kappa\mathcal{C}_{\mbox{sc}}(p_{Y_1,Y_2|X}).
\end{align*}
Since $\mathcal{C}_{\mbox{sc}}(p_{Y_1,Y_2|X})\subseteq\mathcal{C}(p_{Y_1,Y_2|X})\subseteq\mathcal{C}_1(p_{Y_1,Y_2|X})$, the proof is complete.
\end{IEEEproof}

Now we proceed to consider several concrete examples.

\subsubsection{$\mbox{BS-BC}(p_1,p_2)$}

First consider the case where  $p_{Y_1,Y_2|X}$ is a $\mbox{BS-BC}(p_1,p_2)$  with $0\leq p_1\leq p_2<\frac{1}{2}$. Without loss of generality, we shall assume $d_1\leq d_2$.
By Theorem \ref{thm:binary} and Proposition \ref{prop:bsbc} (or by Theorem \ref{thm:binary} and Proposition \ref{prop:symmetry}), if $(\kappa,\mathcal{Q}(w_H,d_1),\mathcal{Q}(w_H,d_2))\in\Gamma$, then
\begin{align}
\mathcal{C}(\mbox{BS-BC}(d_1,d_2))\subseteq\kappa\mathcal{C}(\mbox{BS-BC}(p_1,p_2)).\label{eq:bsbcnec}
\end{align}
On the other hand, the necessary condition implied by the source-channel separation theorem for point-to-point communication systems is
\begin{align}
1-H_b(d_i)\leq\kappa[1-H_b(p_i)],\quad i=1,2.\label{eq:bscsep}
\end{align}
For the special case $\kappa=1$, both (\ref{eq:bsbcnec}) and (\ref{eq:bscsep}) reduce to
\begin{align*}
d_i\geq p_i, \quad i=1,2,
\end{align*}
which is achievable by the uncoded scheme.

In view of Proposition \ref{prop:bsbc} as well as (\ref{eq:derivative0}) and (\ref{eq:derivativehalf}), we have
\begin{align*}
&\phi'_+(0)=\frac{(1-2p_2)\log\Big(\frac{1-p_2}{p_2}\Big)}{(1-2p_1)\log\Big(\frac{1-p_1}{p_1}\Big)},\\
&\phi'_-(C(p_{Y_1|X}))=\frac{(1-2p_2)^2}{(1-2p_1)^2}.
\end{align*}
Hence, it follows from Proposition \ref{prop:nontrivial}  that $\kappa^{\star}>\kappa^{\dagger}$
if
\begin{align}
&\frac{(1-2d_2)\log\Big(\frac{1-d_2}{d_2}\Big)}{(1-2d_1)\log\Big(\frac{1-d_1}{d_1}\Big)}<\frac{(1-2p_2)\log\Big(\frac{1-p_2}{p_2}\Big)}{(1-2p_1)\log\Big(\frac{1-p_1}{p_1}\Big)},\label{eq:satt1}\\
&\frac{(1-2d_2)^2}{(1-2d_1)^2}>\frac{(1-2p_2)^2}{(1-2p_1)^2}.\label{eq:satt2}
\end{align}
For example, (\ref{eq:satt1}) and (\ref{eq:satt2}) are satisfied when $d_1=0.035$, $d_2=0.095$, $p_1=0.15$, and $p_2=0.2$.




\subsubsection{$\mbox{BE-BC}(\epsilon_1,\epsilon_2)$}
Next consider the case where $p_{Y_1,Y_2|X}$ is a $\mbox{BE-BC}(\epsilon_1,\epsilon_2)$  with $0\leq\epsilon_1\leq\epsilon_2<1$. Without loss of generality, we shall assume
$d_1\leq d_2$.
By Proposition \ref{prop:bebc} (or by Proposition \ref{prop:symmetry}),
\begin{align*}
\kappa^{\star}=\min\{\kappa\geq 0: \mathcal{C}(\mbox{BS-BC}(d_1,d_2))\subseteq\kappa\mathcal{C}(\mbox{BE-BC}(\epsilon_1,\epsilon_2))\},
\end{align*}
where the expressions of $\mathcal{C}(\mbox{BS-BC}(d_1,d_2))$ and $\mathcal{C}(\mbox{BE-BC}(\epsilon_1,\epsilon_2))$ can be found in (\ref{eq:bsbcexp1})-(\ref{eq:bsbcexp2}) and (\ref{eq:bebcexp1})-(\ref{eq:bebcexp2}), respectively.
It is clear that, for any $\alpha\in[0,\frac{1}{2}]$, there exists $\beta\in[0,1]$ such that
\begin{align}
&H_b(\alpha *d_1)-H_b(d_1)\leq\kappa^{\star}\beta(1-\epsilon_1),\label{eq:ineq1}\\
&1-H_b(\alpha*d_2)\leq\kappa^{\star}(1-\beta)(1-\epsilon_2),\label{eq:ineq2}
\end{align}
which implies
\begin{align}
\kappa^{\star}\geq\frac{H_b(\alpha *d_1)-H_b(d_1)}{1-\epsilon_1}+\frac{1-H_b(\alpha*d_2)}{1-\epsilon_2}\label{eq:geqany}
\end{align}
for any $\alpha\in[0,\frac{1}{2}]$. Moreover, the equalities must hold in (\ref{eq:ineq1}) and (\ref{eq:ineq2}) for some $\alpha\in[0,\frac{1}{2}]$ and $\beta\in[0,1]$; as a consequence, the equality must hold in (\ref{eq:geqany}) for some $\alpha\in[0,\frac{1}{2}]$.
Therefore, we have
\begin{align}
\kappa^{\star}=\max\limits_{\alpha\in[0,\frac{1}{2}]}\frac{H_b(\alpha *d_1)-H_b(d_1)}{1-\epsilon_1}+\frac{1-H_b(\alpha*d_2)}{1-\epsilon_2},\label{eq:hatkappa}
\end{align}
from which one can readily recover \cite[Theorem 1]{TKS13} by invoking Theorem \ref{thm:binary}. In light of \cite[Lemma 2]{SSC13}, for the optimization problem in (\ref{eq:hatkappa}), the maximum value is not attained at $\alpha=0$ or $\alpha=\frac{1}{2}$ if and only if
\begin{align*}
\frac{(1-2d_2)\log\Big(\frac{1-d_2}{d_2}\Big)}{(1-2d_1)\log\Big(\frac{1-d_1}{d_1}\Big)}<\frac{1-\epsilon_2}{1-\epsilon_1}<\frac{(1-2d_2)^2}{(1-2d_1)^2},
\end{align*}
which gives the necessary and sufficient condition for $\kappa^{\star}>\kappa^{\dagger}$ to hold. The same condition can be obtained through Proposition \ref{prop:trivial} and Proposition \ref{prop:nontrivial}.

\subsubsection{$\mbox{BSC}(p)\&\mbox{BEC}(\epsilon)$}

Finally consider the case where $p_{Y_1,Y_2|X}$ is a $\mbox{BSC}(p)\&\mbox{BEC}(\epsilon)$ with $p\in[0,\frac{1}{2})$ and $\epsilon\in[0,1)$. By Proposition \ref{prop:symmetry},
\begin{align}
&\kappa^{\star}=\min\{\kappa\geq 0: \mathcal{C}(\mbox{BS-BC}(d_1,d_2))\nonumber\\
&\hspace{1.03in}\subseteq\kappa\mathcal{C}(\mbox{BSC}(p)\&\mbox{BEC}(\epsilon))\}.\label{eq:appsym}
\end{align}
Note that
\begin{align*}
\kappa^{\star}\geq\kappa^{\dagger}=\max\Big\{\frac{1-H_b(d_1)}{1-H_b(p)},\frac{1-H_b(d_2)}{1-\epsilon}\Big\}.
\end{align*}

For the case $d_1\leq d_2$, in view of the expression of $\mathcal{C}(\mbox{BSC}(p)\&\mbox{BEC}(\epsilon))$ (see Section \ref{sec:IIIexample}) and the fact that $\frac{\mathrm{d}R_2(\alpha)}{\mathrm{d}R_1(\alpha)}\in[-1,0]$ for $\alpha\in[0,\frac{1}{2}]$, one can readily verify that
\begin{align*}
&\mathcal{C}(\mbox{BS-BC}(d_1,d_2))\subseteq\kappa\mathcal{C}(\mbox{BSC}(p)\&\mbox{BEC}(\epsilon))\\
&\Leftrightarrow \mathcal{C}(\mbox{BS-BC}(d_1,d_2))\subseteq\kappa\mathcal{C}(\mbox{BE-BC}(H_b(p),\epsilon));
\end{align*}
as a consequence,
\begin{align*}
\kappa^{\star}=\max\limits_{\alpha\in[0,\frac{1}{2}]}\frac{H_b(\alpha *d_1)-H_b(d_1)}{1-H_b(p)}+\frac{1-H_b(\alpha*d_2)}{1-\epsilon},
\end{align*}
and we have $\kappa^{\star}>\kappa^{\dagger}$ if and only if
\begin{align*}
\frac{(1-2d_2)\log\Big(\frac{1-d_2}{d_2}\Big)}{(1-2d_1)\log\Big(\frac{1-d_1}{d_1}\Big)}<\frac{1-\epsilon}{1-H_b(p)}<\frac{(1-2d_2)^2}{(1-2d_1)^2}.
\end{align*}


For the case $d_1\geq d_2$, we shall show that
\begin{align}
&\mathcal{C}(\mbox{BS-BC}(d_1,d_2))\subseteq\kappa\mathcal{C}(\mbox{BSC}(p)\&\mbox{BEC}(\epsilon))\nonumber\\
&\Leftrightarrow \mathcal{C}(\mbox{BS-BC}(d_1,d_2))\subseteq\kappa\tilde{\mathcal{C}}(\mbox{BSC}(p)\&\mbox{BEC}(\epsilon)),\label{eq:equivtwo}
\end{align}
where $\tilde{\mathcal{C}}(\mbox{BSC}(p)\&\mbox{BEC}(\epsilon))$ is given by the set\footnote{It follows from \cite[Lemma 2]{WZ73} that $\tilde{\mathcal{C}}(\mbox{BSC}(p)\&\mbox{BEC}(\epsilon))$ is a convex set.} of $(R_1,R_2)\in\mathbb{R}^2_+$ satisfying
\begin{align*}
&R_1\leq 1-H_b(\alpha*p),\\
&R_2\leq(1-\epsilon)H_b(\alpha)
\end{align*}
for some $\epsilon\in[0,\frac{1}{2}]$.
It is easy to see that (\ref{eq:equivtwo}) is true when $\epsilon\in[H_b(p),1)$; moreover,
\begin{align*}
&\mathcal{C}(\mbox{BSC}(p)\&\mbox{BEC}(\epsilon))\\
&=\tilde{\mathcal{C}}(\mbox{BSC}(p)\&\mbox{BEC}(\epsilon))\cap\{(R_1,R_2):R_1+R_2\leq 1-\epsilon\}
\end{align*}
when $\epsilon\in[0,H_b(p))$. Combining this observation with the fact that
\begin{align*}
&\mathcal{C}(\mbox{BS-BC}(d_1,d_2))\subseteq\kappa\tilde{\mathcal{C}}(\mbox{BSC}(p)\&\mbox{BEC}(\epsilon))\\
&\Rightarrow 1-H_b(d_2)\leq\kappa(1-\epsilon)\\
&\stackrel{d_1\geq d_2}{\Rightarrow} \mathcal{C}(\mbox{BS-BC}(d_1,d_2))\subseteq\kappa\{(R_1,R_2):R_1+R_2\leq 1-\epsilon\}
\end{align*}
proves (\ref{eq:equivtwo}). Now we proceed to show that\footnote{This result is not implied by Proposition \ref{prop:trivial}.} $\kappa^{\star}=\kappa^{\dagger}$ if $\kappa^{\dagger}\geq 1$. In view of (\ref{eq:appsym}) and (\ref{eq:equivtwo}), it suffices to show that, if $\kappa^{\dagger}\geq 1$, then
\begin{align}
&1-H_b(\alpha*d_1)\leq\kappa^{\dagger}[1-H_b(\alpha*p)],\label{eq:toprove1}\\
&H_b(\alpha*d_2)-H_b(d_2)\leq\kappa^{\dagger}(1-\epsilon)H_b(\alpha)\label{eq:toprove2}
\end{align}
for any $\alpha\in[0,\frac{1}{2}]$. Note that (\ref{eq:toprove1}) and (\ref{eq:toprove2}) hold when $\alpha=0$ or $\alpha=\frac{1}{2}$. Moreover, $\kappa^{\dagger}\geq 1$ implies $p\geq d_1$. Therefore, an argument similar to that for (\ref{eq:uselater}) can be used here to finish the proof.

\section{The Quadratic Gaussian Case}\label{sec:Gaussian}


Let $\{S(t)\}_{t=1}^\infty$ in System $\Pi$ be an i.i.d. vector Gaussian process, where each $S(t)$ is an $\ell\times 1$ zero-mean Gaussian random vector with positive definite covariance matrix $\Sigma_S$. The following definition is the quadratic Gaussian counterpart of Definition \ref{def:systemA}.

\begin{definition}\label{def:GsystemA}
 Let $\kappa$ be a non-negative number and $\mathcal{D}_i$ be a non-empty compact set of $\ell\times\ell$ positive semi-definite matrices, $i=1,2$. We say $(\kappa,\mathcal{D}_1,\mathcal{D}_2)$ is achievable for System $\Pi$ if, for every $\epsilon>0$, there exist encoding function $f^{(m,n)}:\mathbb{R}^{\ell\times m}\rightarrow\mathcal{X}^n$ and decoding functions $g_i^{(n,m)}:\mathcal{Y}^n_i\rightarrow \mathbb{R}^{\ell\times m}$, $i=1,2$, such that
\begin{align*}
&\frac{n}{m}\leq\kappa+\epsilon,\\
&\min\limits_{D_i\in\mathcal{D}_i}\left\|\frac{1}{m}\sum\limits_{t=1}^m\mathbb{E}[(S(t)-\hat{S}_i(t))(S(t)-\hat{S}_i(t))^T]-D_i\right\|\leq\epsilon,\\
&\hspace{3in}i=1,2.
\end{align*}
The set of all achievable $(\kappa, \mathcal{D}_1,\mathcal{D}_2)$ for System $\Pi$ is denoted by $\Gamma_G$.
\end{definition}

Remark: It is clear that $(\kappa, \mathcal{D}_1,\mathcal{D}_2)\in\Gamma_G$ if and only if $(\kappa, \bar{\mathcal{D}}_1,\bar{\mathcal{D}}_2)\in\Gamma_G$, where
\begin{align*}
\bar{\mathcal{D}}_i=\bigcup\limits_{D_i\in\mathcal{D}_i}\{D'_i:0\preceq D'_i\preceq D_i\},\quad i=1,2.
\end{align*}
Furthermore, to determine whether or not $(\kappa, \bar{\mathcal{D}}_1,\bar{\mathcal{D}}_2)\in\Gamma_G$, there is no loss of generality in setting
$\hat{S}^m_i=\mathbb{E}[S^m|Y^n_i]$, $i=1,2$, for which we have
\begin{align*}
\frac{1}{m}\sum\limits_{t=1}^m\mathbb{E}[(S(t)-\hat{S}_i(t))(S(t)-\hat{S}_i(t))^T]\preceq\Sigma_S,\quad i=1,2.
\end{align*}
Therefore, it suffices to consider those $\mathcal{D}_1$ and $\mathcal{D}_2$ with the property that
\begin{align}
\mathcal{D}_i=\bar{\mathcal{D}}_i\cap\{D:0\preceq D\preceq\Sigma_S\},\quad i=1,2. \label{eq:consideration}
\end{align}
Henceforth we shall implicitly assume that  (\ref{eq:consideration}) is satisfied.


Now we proceed to introduce the corresponding System $\tilde{\Pi}$ in the quadratic Gaussian setting and establish its associated source-channel separation theorem. Let
$\tilde{S}\triangleq(\tilde{S}^T_1,\tilde{S}^T_2)^T$ be an $\tilde{\ell}\times 1$ zero-mean Gaussian random vector with positive definite covariance matrix
$\Sigma_{\tilde{S}}$, where $\tilde{S}_i$ is an $\tilde{\ell}_i\times 1$ random vector, and its covariance matrix is denoted by $\Sigma_{\tilde{S}_i}$, $i=1,2$. Let $\{(\tilde{S}_1(t),\tilde{S}_2(t))\}_{t=1}^\infty$ be i.i.d. copies of $(\tilde{S}_1,\tilde{S}_2)$, and define $\tilde{S}(t)=(\tilde{S}^T_1(t),\tilde{S}^T_2(t))^T$, $t=1,2,\cdots$.

\begin{definition}\label{def:GsystemB}
 Let $\tilde{\kappa}$ be a non-negative number,  $\tilde{\mathcal{D}}_1$ be a non-empty compact subset of  $\{\tilde{D}_1:0\preceq \tilde{D}_1\preceq\Sigma_{\tilde{S}}\}$, and  $\tilde{\mathcal{D}}_2$ be a non-empty compact subset of  $\{\tilde{D}_2:0\preceq \tilde{D}_2\preceq\Sigma_{\tilde{S}_2}\}$.  We say $(\tilde{\kappa},\tilde{\mathcal{D}}_1,\tilde{\mathcal{D}}_2)$ is achievable for System $\tilde{\Pi}$ if, for every $\epsilon>0$, there exist an encoding function $f^{(m,n)}:\mathbb{R}^{\tilde{\ell}_1\times m}\times\mathbb{R}^{\tilde{\ell}_2\times m}\rightarrow\mathcal{X}^n$ as well as decoding functions $g_1^{(n,m)}:\mathcal{Y}^n_1\times\mathbb{R}^{\tilde{\ell}_2\times m}\rightarrow\mathbb{R}^{\tilde{\ell}\times m}$ and $g_2^{(n,m)}:\mathcal{Y}^n_2\rightarrow\mathbb{R}^{\tilde{\ell}_2\times m}$ such that
\begin{align*}
&\frac{n}{m}\leq\tilde{\kappa}+\epsilon,\\
&\min\limits_{\tilde{D}_1\in\tilde{\mathcal{D}}_1}\left\|\sum\limits_{t=1}^m\mathbb{E}[(\tilde{S}(t)-\hat{S}_1(t))(\tilde{S}(t)-\hat{S}_1(t))^T]-\tilde{D}_1\right\|\leq\epsilon,\\
&\min\limits_{\tilde{D}_2\in\tilde{\mathcal{D}}_2}\left\|\sum\limits_{t=1}^m\mathbb{E}[(\tilde{S}_2(t)-\hat{S}_2(t))(\tilde{S}_2(t)-\hat{S}_2(t))^T]-\tilde{D}_2\right\|\leq\epsilon.
\end{align*}
The set of all achievable $(\tilde{\kappa}, \tilde{\mathcal{D}}_1,\tilde{\mathcal{D}}_2)$ for System $\tilde{\Pi}$ is denoted by $\tilde{\Gamma}_G$.
\end{definition}

Remark: Here we allow $f^{(m,n)}$, $g_1^{(n,m)}$, and $g_2^{(n,m)}$ to be non-deterministic functions as long as the Markov chains $(\tilde{S}^m_{1},\tilde{S}^m_{2})\leftrightarrow X^n\leftrightarrow (Y^n_{1},Y^n_{2})$, $\tilde{S}^m_{1}\leftrightarrow(Y^n_{1},\tilde{S}^m_{2})\leftrightarrow\hat{S}^m_{1}$, and $\tilde{S}^m_{2}\leftrightarrow Y^n_{2}\leftrightarrow\hat{S}^m_{2}$ are preserved.

Note that
\begin{align*}
\Sigma_{\tilde{S}}=\left(
                     \begin{array}{cc}
                       \Sigma_{\tilde{S}_1} & \Sigma_{\tilde{S}_1,\tilde{S}_2} \\
                       \Sigma_{\tilde{S}_2,\tilde{S}_1} & \Sigma_{\tilde{S}_2} \\
                     \end{array}
                   \right),
\end{align*}
where $\Sigma_{\tilde{S}_1,\tilde{S}_2}=\mathbb{E}[\tilde{S}_1\tilde{S}^T_2]$ and $\Sigma_{\tilde{S}_2,\tilde{S}_1}=\mathbb{E}[\tilde{S}_2\tilde{S}^T_1]$. Moreover, we write
\begin{align*}
\tilde{D}_1=\left(
              \begin{array}{cc}
                \tilde{D}_{1,1} & \tilde{D}_{1,2} \\
                \tilde{D}_{2,1} & \tilde{D}_{2,2} \\
              \end{array}
            \right)
\end{align*}
for any $\tilde{D}_1\in\tilde{\mathcal{D}}_1$, where $\tilde{D}_{i,i}$ is an $\tilde{\ell}_i\times\tilde{\ell}_i$ matrix, $i=1,2$.
The following source-channel separation theorem is a simple translation of Theorem \ref{thm:systemB} to the quadratic Gaussian setting. Its proof is omitted.

\begin{theorem}\label{thm:Gsep}
$(\tilde{\kappa}, \tilde{\mathcal{D}}_1,\tilde{\mathcal{D}}_2)\in\tilde{\Gamma}_G$ if and only if
$(R_{\tilde{S}_1|\tilde{S}_2}(\tilde{\mathcal{D}}_1),R_{\tilde{S}_2}(\tilde{\mathcal{D}}_2))\in\tilde{\kappa}\mathcal{C}_{1}(p_{Y_1,Y_2|X})$, where
\begin{align*}
&R_{\tilde{S}_1|\tilde{S}_2}(\tilde{\mathcal{D}}_1)=\min\limits_{\tilde{D}_1\in\tilde{\mathcal{D}}_1}\frac{1}{2}\log\Big(\frac{|\Sigma_{\tilde{S}_1}-\Sigma_{\tilde{S}_1,\tilde{S}_2}\Sigma^{-1}_{\tilde{S}_2}\Sigma_{\tilde{S}_2,\tilde{S}_1}|}{|\tilde{D}_{1,1}-K\tilde{D}_{2,1}|}\Big),\\
&R_{\tilde{S}_2}(\tilde{\mathcal{D}}_2)=\min\limits_{\tilde{D}_2\in\tilde{\mathcal{D}}_2}\frac{1}{2}\log\Big(\frac{|\Sigma_{\tilde{S}_2}|}{|\tilde{D}_2|}\Big)
\end{align*}
with $K$ being any solution\footnote{If $\tilde{D}_{2,2}$ is invertible, then $K=\tilde{D}_{1,2}\tilde{D}^{-1}_{2,2}$.} of $K\tilde{D}_{2,2}=\tilde{D}_{1,2}$.
\end{theorem}

Remark: It can be verified that
\begin{align*}
&R_{\tilde{S}_1|\tilde{S}_2}(\tilde{\mathcal{D}}_1)=\min\limits_{p_{\hat{S}_1|\tilde{S}}:\mathbb{E}[(\tilde{S}-\hat{S}_1)(\tilde{S}-\hat{S}_1)^T]\in\tilde{\mathcal{D}}_1}I(\tilde{S}_1;\hat{S}_1|\tilde{S}_2),\\
&R_{\tilde{S}_2}(\tilde{\mathcal{D}}_2)=\min\limits_{p_{\hat{S}_2|\tilde{S}_2}:\mathbb{E}[(\tilde{S}_2-\hat{S}_2)(\tilde{S}_2-\hat{S}_2)^T]\in\tilde{\mathcal{D}}_2}I(\tilde{S}_2;\hat{S}_2),
\end{align*}
which highlights the similarity between Theorem \ref{thm:systemB} and Theorem \ref{thm:Gsep}.


Again, in the quadratic Gaussian setting,  the source-channel separation theorem for System $\tilde{\Pi}$ can be leveraged to derive a necessary condition for System $\Pi$.
For any $D_i\in\mathcal{D}_i$, $i=1,2$, let $\mathcal{R}_1(\Sigma_S,D_1,D_2)$ denote the convex closure of the set of $(R_1,R_2)\in\mathbb{R}^2_+$ satisfying
\begin{align*}
&R_1\leq\frac{1}{2}\log\Big(\frac{|\Sigma_S||D_1+\Sigma_Z|}{|D_1||\Sigma_S+\Sigma_Z|}\Big),\\
&R_2\leq\frac{1}{2}\log\Big(\frac{|\Sigma_S+\Sigma_Z|}{|D_2+\Sigma_Z|}\Big)
\end{align*}
for some $\Sigma_Z\succ 0$, and let $\mathcal{R}_2(\Sigma_S,D_1,D_2)$ denote the convex closure of the set of $(R_1,R_2)\in\mathbb{R}^2_+$ satisfying
\begin{align*}
&R_1\leq\frac{1}{2}\log\Big(\frac{|\Sigma_S+\Sigma_Z|}{|D_1+\Sigma_Z|}\Big),\\
&R_2\leq\frac{1}{2}\log\Big(\frac{|\Sigma_S||D_2+\Sigma_Z|}{|D_2||\Sigma_S+\Sigma_Z|}\Big)
\end{align*}
for some $\Sigma_Z\succ 0$. By setting $\Sigma_U=\Sigma_S(\Sigma_S+\Sigma_Z)^{-1}\Sigma_S$, we can write $\mathcal{R}_1(\Sigma_S,D_1,D_2)$ equivalently as the convex hull of the set of $(R_1,R_2)\in\mathbb{R}^2_+$ such that
\begin{align*}
&R_1\leq\frac{1}{2}\log\Big(\frac{|\Sigma_U\Sigma^{-1}_SD_1+\Sigma_S-\Sigma_U|}{|D_1|}\Big),\\
&R_2\leq\frac{1}{2}\log\Big(\frac{|\Sigma_S|}{|\Sigma_U\Sigma^{-1}_SD_2+\Sigma_S-\Sigma_U|}\Big)
\end{align*}
for some $\Sigma_U$ satisfying $0\preceq\Sigma_U\preceq\Sigma_S$; similarly,  $\mathcal{R}_2(\Sigma_S,D_1,D_2)$ can be written equivalently as the convex hull of the set of $(R_1,R_2)\in\mathbb{R}^2_+$ such that
\begin{align*}
&R_1\leq\frac{1}{2}\log\Big(\frac{|\Sigma_S|}{|\Sigma_U\Sigma^{-1}_SD_1+\Sigma_S-\Sigma_U|}\Big),\\
&R_2\leq\frac{1}{2}\log\Big(\frac{|\Sigma_U\Sigma^{-1}_SD_2+\Sigma_S-\Sigma_U|}{|D_2|}\Big)
\end{align*}
for some $\Sigma_U$ satisfying $0\preceq\Sigma_U\preceq\Sigma_S$.

  Let $S$ be an $\ell\times 1$ zero-mean Gaussian random vector with positive definite covariance matrix $\Sigma_S$. Recall the definition of $\mathcal{R}_i(p_{S,\hat{S}_1,\hat{S}_2})$, $i=1,2$ in  Section \ref{sec:application}.  The following result provides a connection between $\mathcal{R}_i(\Sigma_S,D_1,D_2)$ and $\mathcal{R}_i(p_{S,\hat{S}_1,\hat{S}_2})$, $i=1,2$.

\begin{proposition}\label{prop:justifyG}
If $\mathbb{E}[(S-\hat{S}_i)(S-\hat{S}_i)^T]=D_i\in\mathcal{D}_i$, $i=1,2$, then
\begin{align}
\mathcal{R}_i(p_{S,\hat{S}_1,\hat{S}_2})\supseteq\mathcal{R}_i(\Sigma_S,D_1,D_2),\quad i=1,2.\label{eq:Gjust1}
\end{align}
Moreover, if $S-\hat{S}_i$ and $\hat{S}_i$ are independent zero-mean Gaussian random vectors with covariance matrices $D_i$ and $\Sigma_S-D_i$, respectively, $i=1,2$, where $0\preceq D_1\preceq D_2\preceq\Sigma_S$, then
\begin{align}
&\mathcal{R}_1(p_{S,\hat{S}_1,\hat{S}_2})=\mathcal{R}_1(\Sigma_S,D_1,D_2),\label{eq:Gjust2}\\
&\mathcal{R}_2(p_{S,\hat{S}_1,\hat{S}_2})\subseteq\Big\{(R_1,R_2)\in\mathbb{R}^2_+:R_2\leq\frac{1}{2}\log\Big(\frac{|\Sigma_S|}{|D_2|}\Big),\nonumber\\
&\hspace{1.5in}R_1+R_2\leq\frac{1}{2}\log\Big(\frac{|\Sigma_S|}{|D_1|}\Big)\Big\}.\label{eq:Gjust3}
\end{align}
\end{proposition}
\begin{IEEEproof}
By symmetry, it suffices to prove (\ref{eq:Gjust1}) for $i=1$. Given any $\Sigma_U$ satisfying $0\preceq\Sigma_U\preceq\Sigma_S$, we can find $U$ jointly distributed with $S$ such that $U$ and $S-U$ are independent zero-mean Gaussian random vectors with covariance matrices $\Sigma_U$ and $\Sigma_S-\Sigma_U$, respectively. Note that for any $(\hat{S}_1,\hat{S}_2)$ jointly distributed with such $(U,S)$ subject to the constraints that  $\mathbb{E}[(S-\hat{S}_i)(S-\hat{S}_i)^T]=D_i\in\mathcal{D}_i$, $i=1,2$, and that $U\leftrightarrow S\leftrightarrow (\hat{S}_1,\hat{S}_2)$ form a Markov chain,  we have
\begin{align}
&I(S;\hat{S}_1|U)\geq\frac{1}{2}\log\Big(\frac{|\Sigma_U\Sigma^{-1}_SD_1+\Sigma_S-\Sigma_U|}{|D_1|}\Big),\label{eq:Ueq1}\\
&I(U;\hat{S}_2)\geq\frac{1}{2}\log\Big(\frac{|\Sigma_S|}{|\Sigma_U\Sigma^{-1}_SD_2+\Sigma_S-\Sigma_U|}\Big),\label{eq:Ueq2}
\end{align}
where the equalities in (\ref{eq:Ueq1}) and (\ref{eq:Ueq2}) hold when $S-\hat{S}_i$ and $\hat{S}_i$ are independent zero-mean Gaussian random vectors with covariance matrices $D_i$ and $\Sigma_S-D_i$, respectively, $i=1,2$. Now the desired result follows by the convexity of $\mathcal{R}_1(p_{S,\hat{S}_1,\hat{S}_2})$.

To prove (\ref{eq:Gjust2}), it suffices to consider the non-degenerate case $0\prec D_1\preceq D_2\prec\Sigma_S$; the general case $0\preceq D_1\preceq D_2\preceq\Sigma_S$ can be proved via a simple limiting argument. Let $O_i$ be a zero-mean Gaussian random vector, independent of $(U,S)$, with covariance matrix $\Sigma_{O_i}=(D^{-1}_i-\Sigma^{-1}_S)^{-1}$, $i=1,2$. It is clear that
\begin{align*}
&I(S;\hat{S}_1|U)=I(S;S+O_1|U),\\
&I(U;\hat{S}_2)=I(U;S+O_2).
\end{align*}
For any $\lambda\in[0,1]$,
\begin{align}
&\max\limits_{(R_1,R_2)\in\mathcal{R}_1(p_{S,\hat{S}_1,\hat{S}_2})}\lambda R_1+(1-\lambda)R_2\nonumber\\
&=\max_{p_{U|S}}\lambda I(S;\hat{S}_1|U)+(1-\lambda)I(U;\hat{S}_2)\nonumber\\
&=\max_{p_{U|S}}\lambda I(S;S+O_1|U)+(1-\lambda)I(U;S+O_2)\nonumber\\
&=\max\limits_{0\preceq\Sigma_U\preceq\Sigma_S}\frac{\lambda}{2}\log\Big(\frac{|\Sigma_S-\Sigma_U+\Sigma_{O_1}|}{|\Sigma_{O_1}|}\Big)\nonumber\\
&\hspace{0.7in}+\frac{1-\lambda}{2}\log\Big(\frac{|\Sigma_S+\Sigma_{O_2}|}{|\Sigma_S-\Sigma_U+\Sigma_{O_2}|}\Big)\label{eq:lvextr}\\
&=\max\limits_{0\preceq\Sigma_U\preceq\Sigma_S}\frac{\lambda}{2}\log\Big(\frac{|\Sigma_U\Sigma^{-1}_SD_1+\Sigma_S-\Sigma_U|}{|D_1|}\Big)\nonumber\\
&\hspace{0.7in}+\frac{1-\lambda}{2}\log\Big(\frac{|\Sigma_S|}{|\Sigma_U\Sigma^{-1}_SD_2+\Sigma_S-\Sigma_U|}\Big)\nonumber\\
&=\max\limits_{(R_1,R_2)\in\mathcal{R}_1(\Sigma_S,D_1,D_2)}\lambda R_1+(1-\lambda)R_2,\nonumber
\end{align}
where (\ref{eq:lvextr}) is due to the conditional version of \cite[Corollary 4]{LV07}. This together with the convexity of $\mathcal{R}_1(p_{S,\hat{S}_1,\hat{S}_2})$ and $\mathcal{R}_1(\Sigma_S,D_1,D_2)$ proves (\ref{eq:Gjust2}). It can be verified that
\begin{align*}
I(S;\hat{S}_2|U)&\leq I(S;\hat{S}_2)\\
&\leq\frac{1}{2}\log\Big(\frac{|\Sigma_S|}{|D_2|}\Big)
\end{align*}
and
\begin{align*}
I(U;\hat{S}_1)+I(S;\hat{S}_2|U)&\leq I(U;\hat{S}_1)+I(S;\hat{S}_1|U)\\
&=I(S;\hat{S}_1)\\
&=\frac{1}{2}\log\Big(\frac{|\Sigma_S|}{|D_1|}\Big),
\end{align*}
from which (\ref{eq:Gjust3}) follows immediately.
\end{IEEEproof}

\begin{theorem}\label{thm:Gaussian}
For any $(\kappa,\mathcal{D}_1,\mathcal{D}_2)\in\Gamma_G$, there exist $D_i\in\mathcal{D}_i$, $i=1,2$,
such that
\begin{align}
\mathcal{R}_i(\Sigma_S,D_1,D_2)\subseteq\kappa\mathcal{C}_{i}(p_{Y_1,Y_2|X}),\quad i=1,2.\label{eq:imp}
\end{align}
\end{theorem}
\begin{IEEEproof}
By symmetry, it suffices to prove (\ref{eq:imp}) for $i=1$. Let $\{Z(t)\}_{t=1}^\infty$ be an i.i.d. vector Gaussian process, independent of $\{S(t)\}_{t=1}^\infty$, where each $Z(t)$ is an $\ell\times 1$ zero-mean Gaussian random vector with positive definite covariance matrix $\Sigma_Z$. Define $\tilde{S}_1(t)=S(t)$ and $\tilde{S}_2(t)=S(t)+Z(t)$ for $t=1,2,\cdots$.
 Now consider an arbitrary tuple $(\kappa,\mathcal{D}_1,\mathcal{D}_2)\in\Gamma_G$. Given any $\epsilon>0$, according to Definition \ref{def:GsystemA}, there exist encoding function $f^{(m,n)}:\mathbb{R}^{\ell\times m}\rightarrow\mathcal{X}^n$ and decoding functions $g_i^{(n,m)}:\mathcal{Y}^n_i\rightarrow\mathbb{R}^{\ell\times m}$, $i=1,2$, satisfying\footnote{We have denoted $\hat{S}_i(t)$ by $\hat{S}^{(\epsilon)}_i(t)$ to stress its dependence on $\epsilon$}
 \begin{align*}
&\frac{n}{m}\leq\kappa+\epsilon,\\
&\min\limits_{D_i\in\mathcal{D}_i}\left\|\frac{1}{m}\sum\limits_{t=1}^m\mathbb{E}[(S(t)-\hat{S}^{(\epsilon)}_i(t))(S(t)-\hat{S}^{(\epsilon)}_i(t))^T]-D_i\right\|\\
&\leq\epsilon,\quad i=1,2.
 \end{align*}
 Therefore, one can find a sequence $\epsilon_1, \epsilon_2, \cdots$ converging to zero such that
 \begin{align}
 \lim\limits_{k\rightarrow\infty}\frac{1}{m}\sum\limits_{t=1}^m\mathbb{E}[(S(t)-\hat{S}^{(\epsilon_k)}_i(t))(S(t)-\hat{S}^{(\epsilon_k)}_i(t))^T]=D_i\label{eq:usedlater}
 \end{align}
 for some $D_i\in\mathcal{D}_i$, $i=1,2$. Note that
 \begin{align*}
 &\lim\limits_{k\rightarrow\infty}\frac{1}{m}\sum\limits_{t=1}^m\mathbb{E}[(\tilde{S}_1(t)-\hat{S}^{(\epsilon_k)}_1(t))(\tilde{S}_1(t)-\hat{S}^{(\epsilon_k)}_1(t))^T]\\
 &=\lim\limits_{k\rightarrow\infty}\frac{1}{m}\sum\limits_{t=1}^m\mathbb{E}[(\tilde{S}_1(t)-\hat{S}^{(\epsilon_k)}_1(t))(\tilde{S}_2(t)-\hat{S}^{(\epsilon_k)}_1(t))^T]\\
 &=\lim\limits_{k\rightarrow\infty}\frac{1}{m}\sum\limits_{t=1}^m\mathbb{E}[(\tilde{S}_2(t)-\hat{S}^{(\epsilon_k)}_1(t))(\tilde{S}_1(t)-\hat{S}^{(\epsilon_k)}_1(t))^T]\\
 &=D_1,\\
 &\lim\limits_{k\rightarrow\infty}\frac{1}{m}\sum\limits_{t=1}^m\mathbb{E}[(\tilde{S}_2(t)-\hat{S}^{(\epsilon_k)}_1(t))(\tilde{S}_2(t)-\hat{S}^{(\epsilon_k)}_1(t))^T] \\
 &=D_1+\Sigma_{Z},\\ 
 &\lim\limits_{k\rightarrow\infty}\frac{1}{m}\sum\limits_{t=1}^m\mathbb{E}[(\tilde{S}_2(t)-\hat{S}^{(\epsilon_k)}_2(t))(\tilde{S}_2(t)-\hat{S}^{(\epsilon_k)}_2(t))^T]\\
 &=\tilde{D}_2\triangleq D_2+\Sigma_Z.
 \end{align*}
 As a consequence, we must have $(\kappa,\{\tilde{D}_1\},\{\tilde{D}_2\})\in\tilde{\Gamma}_G$, where
 \begin{align*}
 \tilde{D}_1=\left(
                                                                                               \begin{array}{cc}
                                                                                                 D_1 & D_1 \\
                                                                                                 D_1 & D_1+\Sigma_{Z} \\
                                                                                               \end{array}
                                                                                             \right).
 \end{align*}
 It then follows from Theorem \ref{thm:Gsep} that
\begin{align*}
&\Big(\frac{1}{2}\log\Big(\frac{|\Sigma_S-\Sigma_S(\Sigma_S+\Sigma_Z)^{-1}\Sigma_S|}{|D_1-D_1(D_1+\Sigma_Z)^{-1}D_1|}\Big),\frac{1}{2}\log\Big(\frac{|\Sigma_S+\Sigma_Z|}{|D_2+\Sigma_Z|}\Big)\Big)\\
&\in\kappa\mathcal{C}_{1}(p_{Y_1,Y_2|X}).
\end{align*}
Here one can fix $(D_1,D_2)$ and choose the positive definite covariance matrix $\Sigma_Z$ arbitrarily; moreover, it can be verified that
\begin{align*}
  \frac{|\Sigma_S-\Sigma_S(\Sigma_S+\Sigma_Z)^{-1}\Sigma_S|}{|D_1-D_1(D_1+\Sigma_Z)^{-1}D_1|} &=\frac{|D^{-1}_1+\Sigma^{-1}_Z|}{|\Sigma^{-1}_S+\Sigma^{-1}_Z|}\\
  &=\frac{|\Sigma_S||D_1+\Sigma_Z|}{|D_1||\Sigma_S+\Sigma_Z|}.
\end{align*}
This completes the proof of Theorem \ref{thm:Gaussian}.
\end{IEEEproof}



Note that $\mathcal{R}_1(\Sigma_S,D_1,D_2)$ coincides with the capacity region of vector Gaussian broadcast channel with covariance power constraint $\Sigma_S$ and noise covariances $\Delta_i\triangleq(D^{-1}_i-\Sigma^{-1}_S)^{-1}$, $i=1,2$,  when $0\prec D_1\preceq D_2\prec\Sigma_S$. For this reason, we shall denote $\mathcal{R}_1(\Sigma_S,D_1,D_2)$ alternatively by $\mathcal{C}(\mbox{G-BC}(\Sigma_S,\Delta_1,\Delta_2))$ (even when $\Delta_1$ and $\Delta_2$ are not well-defined). One can obtain the following refined necessary condition for the case where $p_{Y_1,Y_2|X}$ is a scalar Gaussian broadcast channel.
\begin{theorem}\label{thm:Gchannel}
If $p_{Y_1,Y_2|X}$ is a $\mbox{G-BC}(P,N_1,N_2)$ with $0<N_1\leq N_2$, then, for any $(\kappa,\mathcal{D}_1,\mathcal{D}_2)\in\Gamma_G$, there exist $D_i\in\mathcal{D}_i$, $i=1,2$, with $D_1\preceq D_2$
such that
\begin{align*}
\mathcal{C}(\mbox{G-BC}(\Sigma_S,\Delta_1,\Delta_2))\subseteq\kappa\mathcal{C}(\mbox{G-BC}(P, N_1,N_2)).
\end{align*}
\end{theorem}
\begin{IEEEproof}
According to the remark after Definition \ref{def:GsystemA}, there is no loss of generality in setting $\hat{S}^m_i=\mathbb{E}[S^m|Y^n_i]$, $i=1,2$. As a consequence, in (\ref{eq:usedlater}) we must have $D_1\preceq D_2$
if $p_{Y_2|X}$ is degraded with respect to $p_{Y_1|X}$. Now one can readily adapt the proof of Theorem \ref{thm:Gaussian} to the current setting to show that, for any $(\kappa,\mathcal{D}_1,\mathcal{D}_2)\in\Gamma_G$, there exist $D_i\in\mathcal{D}_i$, $i=1,2$, with $D_1\preceq D_2$, such that
\begin{align}
&\mathcal{R}_i(\Sigma_S,D_1,D_2)\subseteq\kappa\mathcal{C}_i(\mbox{G-BC}(P,N_1,N_2)),\quad i=1,2.\label{eq:GCimp}
\end{align}
It follows from Proposition \ref{prop:gbc} that $\mathcal{C}_1(\mbox{G-BC}(P,N_1,N_2))=\mathcal{C}(\mbox{G-BC}(P,N_1,N_2))$, and $\mathcal{C}_2(\mbox{G-BC}(P,N_1,N_2))$ is given by the set of $(R_1,R_2)\in\mathbb{R}^2_+$ satisfying
\begin{align*}
&R_2\leq\frac{1}{2}\log\Big(\frac{P+N_2}{N_2}\Big),\\
&R_1+R_2\leq\frac{1}{2}\log\Big(\frac{P+N_1}{N_1}\Big).
\end{align*}
Note that $\mathcal{R}_1(\Sigma_S,D_1,D_2)\subseteq\kappa\mathcal{C}_1(\mbox{G-BC}(P,N_1,N_2))$ implies
\begin{align*}
\frac{1}{2}\log\Big(\frac{|\Sigma_S|}{|D_i|}\Big)\leq\frac{\kappa}{2}\log\Big(\frac{P+N_i}{N_i}\Big),\quad i=1,2.
\end{align*}
Moreover, in view of (\ref{eq:Gjust1}) and (\ref{eq:Gjust3}) in Proposition \ref{prop:justifyG}, we have
\begin{align*}
&\mathcal{R}_2(\Sigma_S,D_1,D_2)\subseteq\Big\{(R_1,R_2)\in\mathbb{R}^2_+:R_2\leq\frac{1}{2}\log\Big(\frac{|\Sigma_S|}{|D_2|}\Big),\\
&\hspace{1.8in}R_1+R_2\leq\frac{1}{2}\log\Big(\frac{|\Sigma_S|}{|D_1|}\Big)\Big\}.
\end{align*}
Therefore,
\begin{align*}
&\mathcal{R}_1(\Sigma_S,D_1,D_2)\subseteq\kappa\mathcal{C}_1(\mbox{G-BC}(P,N_1,N_2))\\
&\Rightarrow\mathcal{R}_2(\Sigma_S,D_1,D_2)\subseteq\kappa\mathcal{C}_2(\mbox{G-BC}(P,N_1,N_2))
\end{align*}
 when $0\preceq D_1\preceq D_2\preceq\Sigma_{S}$. This completes the proof of Theorem \ref{thm:Gchannel}.
\end{IEEEproof}

For the case $0\preceq D_1\preceq D_2\preceq\Sigma_S$, one can show by leveraging Proposition \ref{prop:justifyG} that (\ref{eq:GCimp}) is equivalent to the existence of $(\hat{S}_1,\hat{S}_2)$ with $\mathbb{E}[(S-\hat{S}_i)(S-\hat{S}_i)^T]=D_i\in\mathcal{D}_i$, $i=1,2$, such that
\begin{align*}
\mathcal{R}_i(p_{S,\hat{S}_1,\hat{S_2}})\subseteq\kappa\mathcal{C}_i(\mbox{G-BC}(P,N_1,N_2)),\quad i=1,2;
\end{align*}
in fact, there is no loss of generality in assuming that $S-\hat{S}_i$ and $\hat{S}_i$ are independent zero-mean Gaussian random vectors with covariance matrices $D_i$ and $\Sigma_S-D_i$, respectively, $i=1,2$. Note that $U$ is not restricted to the form $U=S+Z$ (or equivalently $U=\mathbb{E}[S|S+Z]$) in the definition of $\mathcal{R}_i(p_{S,\hat{S}_1,\hat{S_2}})$, $i=1,2$,  where $Z$ is a zero-mean Gaussian random vector independent of $S$. Therefore, removing this restriction does not lead to a stronger necessary condition. This provides a certain justification for the choice of the auxiliary random variable in \cite{RFZ06}.

With no essential loss of generality, henceforth we focus on the non-degenerate case $\kappa>0$. Define
\begin{align*}
&P^{\star}=\min\{P\geq 0:\mathcal{C}(\mbox{G-BC}(\Sigma_S,\Delta_1,\Delta_2))\\
&\hspace{1.1in}\subseteq\kappa\mathcal{C}(\mbox{G-BC}(P, N_1,N_2))\}.
\end{align*}
It is clear that, for any $\Sigma_Z\succ 0$, there exists $\beta\in[0,1]$ such that
\begin{align*}
&\frac{1}{2}\log\Big(\frac{|\Sigma_S||D_1+\Sigma_Z|}{|D_1||\Sigma_S+\Sigma_Z|}\Big)\leq\frac{\kappa}{2}\log\Big(\frac{\beta P^{\star}+N_1}{N_1}\Big), \\
&\frac{1}{2}\log\Big(\frac{|\Sigma_S+\Sigma_Z|}{|D_2+\Sigma_Z|}\Big)\leq\frac{\kappa}{2}\log\Big(\frac{P^{\star}+N_2}{\beta P^{\star}+N_2}\Big),
\end{align*}
which can be rewritten as
\begin{align*}
&\beta P^{\star}\geq N_1\Big(\frac{|\Sigma_S||D_1+\Sigma_Z|}{|D_1||\Sigma_S+\Sigma_Z|}\Big)^{\frac{1}{\kappa}}-N_1,\\
&\beta P^{\star}\leq (P^{\star}+N_2)\Big(\frac{|D_2+\Sigma_Z|}{|\Sigma_S+\Sigma_Z|}\Big)^{\frac{1}{\kappa}}-N_2.
\end{align*}
Hence, for any $\Sigma_Z\succ 0$, we have
\begin{align*}
&(P^{\star}+N_2)\Big(\frac{|D_2+\Sigma_Z|}{|\Sigma_S+\Sigma_Z|}\Big)^{\frac{1}{\kappa}}-N_2\\
&\geq N_1\Big(\frac{|\Sigma_S||D_1+\Sigma_Z|}{|D_1||\Sigma_S+\Sigma_Z|}\Big)^{\frac{1}{\kappa}}-N_1,
\end{align*}
i.e.,
\begin{align}
P^{\star}&\geq N_1\Big(\frac{|\Sigma_S||D_1+\Sigma_Z|}{|D_1||D_2+\Sigma_Z|}\Big)^{\frac{1}{\kappa}}\nonumber\\
&\quad+(N_2-N_1)\Big(\frac{|\Sigma_S+\Sigma_Z|}{|D_2+\Sigma_Z|}\Big)^{\frac{1}{\kappa}}-N_2.\label{eq:Gcom1}
\end{align}
Moreover, there must exist some $\beta\in[0,1]$ and a sequence of positive definite matrices $\Sigma^{(k)}_Z$, $k=1,2,\cdots$, such that
\begin{align*}
&\lim\limits_{k\rightarrow\infty}\frac{1}{2}\log\Big(\frac{|\Sigma_S||D_1+\Sigma^{(k)}_Z|}{|D_1||\Sigma_S+\Sigma^{(k)}_Z|}\Big)=\frac{\kappa}{2}\log\Big(\frac{\beta P^{\star}+N_1}{N_1}\Big), \\
&\lim\limits_{k\rightarrow\infty}\frac{1}{2}\log\Big(\frac{|\Sigma_S+\Sigma^{(k)}_Z|}{|D_2+\Sigma^{(k)}_Z|}\Big)\Big)=\frac{\kappa}{2}\log\Big(\frac{P^{\star}+N_2}{\beta P^{\star}+N_2}\Big),
\end{align*}
which implies
\begin{align}
&P^{\star}=\lim\limits_{k\rightarrow\infty} N_1\Big(\frac{|\Sigma_S||D_1+\Sigma^{(k)}_Z|}{|D_1||D_2+\Sigma^{(k)}_Z|}\Big)^{\frac{1}{\kappa}}\nonumber\\
&\hspace{0.7in}+(N_2-N_1)\Big(\frac{|\Sigma_S+\Sigma^{(k)}_Z|}{|D_2+\Sigma^{(k)}_Z|}\Big)^{\frac{1}{\kappa}}-N_2.\label{eq:Gcom2}
\end{align}
Combining (\ref{eq:Gcom1}) and (\ref{eq:Gcom2}) gives
\begin{align}
&P^{\star}=\sup\limits_{\Sigma_Z\succ 0} N_1\Big(\frac{|\Sigma_S||D_1+\Sigma_Z|}{|D_1||D_2+\Sigma_Z|}\Big)^{\frac{1}{\kappa}}\nonumber\\
&\hspace{0.65in}+(N_2-N_1)\Big(\frac{|\Sigma_S+\Sigma_Z|}{|D_2+\Sigma_Z|}\Big)^{\frac{1}{\kappa}}-N_2.\label{eq:G2unre}
\end{align}
Therefore, by Theorem \ref{thm:Gchannel},  if $(\kappa,\mathcal{D}_1,\mathcal{D}_2)\in\Gamma_G$, then
\begin{align}
&P\geq\inf\limits_{D_1,D_2}\sup\limits_{\Sigma_Z\succ 0} N_1\Big(\frac{|\Sigma_S||D_1+\Sigma_Z|}{|D_1||D_2+\Sigma_Z|}\Big)^{\frac{1}{\kappa}}\nonumber\\
&\hspace{0.95in}+(N_2-N_1)\Big(\frac{|\Sigma_S+\Sigma_Z|}{|D_2+\Sigma_Z|}\Big)^{\frac{1}{\kappa}}-N_2,\label{eq:generalGaussian}
\end{align}
where the infimum is over $D_1$ and $D_2$ subject to the constraints $D_i\in\mathcal{D}_i$, $i=1,2$, and $D_1\preceq D_2$. For the case where $\mathcal{D}_i=\{D_i:0\preceq D_i\preceq\Theta_i\}$, $i=1,2$, for some $\Theta_1$ and $\Theta_2$ satisfying $0\prec\Theta_1\preceq\Theta_2\preceq\Sigma_S$, we can simplify (\ref{eq:generalGaussian}) to
\begin{align*}
&P\geq\sup\limits_{\Sigma_Z\succ 0} N_1\Big(\frac{|\Sigma_S||\Theta_1+\Sigma_Z|}{|\Theta_1||\Theta_2+\Sigma_Z|}\Big)^{\frac{1}{\kappa}}\nonumber\\
&\hspace{0.65in}+(N_2-N_1)\Big(\frac{|\Sigma_S+\Sigma_Z|}{|\Theta_2+\Sigma_Z|}\Big)^{\frac{1}{\kappa}}-N_2,
\end{align*}
from which one can readily recover \cite[Theorem 1]{RFZ06} by setting $\ell=1$.

Now partition $S(t)$ to the form $S(t)=(S^T_1(t),S^T_2(t))^T$, $t=1,2,\cdots$, where each $S_i(t)$ is an $\ell_i\times 1$ zero-mean Gaussian random vector with positive definite covariance matrix $\Sigma_{S_i}$, $i=1,2$. We require that  $\{S_i(t)\}_{t=1}^\infty$ be reconstructed at receiver $i$ subject to positive definite covariance distortion constraint $\Lambda_i$, $i=1,2$. This corresponds to the case where $\mathcal{D}_i=\mathcal{D}_i(\Lambda_i)\triangleq\{D_i:0\preceq D_i\preceq\Sigma_S, D_{i,i}\preceq\Lambda_i\}$  with $D_i$ partitioned to the form
\begin{align*}
D_i=\left(
           \begin{array}{cc}
             D_{i,1} & \# \\
             \# & D_{i,2} \\
           \end{array}
         \right), \quad i=1,2.
\end{align*}
Therefore, the lower bound in (\ref{eq:generalGaussian}) is also applicable here.
By restricting $\Sigma_Z$ to a special block diagonal form\footnote{Here $I$ is an $\ell_1\times\ell_1$ identity matrix}
\begin{align*}
\Sigma_Z=\left(
           \begin{array}{cc}
             \lambda I & 0 \\
             0 & \Sigma_{Z_2} \\
           \end{array}
         \right),
\end{align*}
one can deduce from (\ref{eq:generalGaussian})
\begin{align}
P&\geq\inf\limits_{D_1,D_2}\sup\limits_{\Sigma_{Z_2}\succ 0}\lim\limits_{\lambda\rightarrow\infty}N_1\Big(\frac{|\Sigma_S||D_1+\Sigma_Z|}{|D_1||D_2+\Sigma_Z|}\Big)^{\frac{1}{\kappa}}\nonumber\\
&\hspace{1.2in}+(N_2-N_1)\Big(\frac{|\Sigma_S+\Sigma_Z|}{|D_2+\Sigma_Z|}\Big)^{\frac{1}{\kappa}}-N_2\nonumber\\
&=\inf\limits_{D_1,D_2}\sup\limits_{\Sigma_{Z_2}\succ 0}N_1\Big(\frac{|\Sigma_S||D_{1,2}+\Sigma_{Z_2}|}{|D_1||D_{2,2}+\Sigma_{Z_2}|}\Big)^{\frac{1}{\kappa}}\nonumber\\
&\hspace{0.7in}+(N_2-N_1)\Big(\frac{|\Sigma_{S_2}+\Sigma_{Z_2}|}{|D_{2,2}+\Sigma_{Z_2}|}\Big)^{\frac{1}{\kappa}}-N_2,\label{eq:Pbound}
\end{align}
where the infimum is over $D_1$ and $D_2$ subject to the constraints $D_i\in\mathcal{D}_i(\Lambda_i)$, $i=1,2$, and $D_1\preceq D_2$. This potentially weakened lower bound, when specialized to the case $\kappa=1$, is at least as tight as \cite[Theorem 1]{SCT15}.
Note that, for any $D_i\in\mathcal{D}_i(\Lambda_i)$, $i=1,2$, and any positive definite matrix $\Sigma_Z$ partitioned to the form
\begin{align}
\Sigma_Z=\left(
           \begin{array}{cc}
             \Sigma_{Z_1} & \# \\
             \# & \Sigma_{Z_2} \\
           \end{array}
         \right),\label{eq:Zpar}
\end{align}
we have
\begin{align}
\frac{|\Sigma_S||D_{1,2}+\Sigma_{Z_2}|}{|D_1||D_{2,2}+\Sigma_{Z_2}|}&\geq\frac{|\Sigma_S+\Sigma_Z||D_{1,2}+\Sigma_{Z_2}|}{|D_1+\Sigma_Z||D_{2,2}+\Sigma_{Z_2}|}\nonumber\\
&\geq\frac{|\Sigma_S+\Sigma_Z|}{|D_{1,1}+\Sigma_{Z_1}||D_{2,2}+\Sigma_{Z_2}|}\nonumber\\
&\geq\frac{|\Sigma_S+\Sigma_Z|}{|\Lambda_1+\Sigma_{Z_1}||\Lambda_2+\Sigma_{Z_2}|}\label{eq:Zsub1}
\end{align}
and
\begin{align}
\frac{|\Sigma_{S_2}+\Sigma_{Z_2}|}{|D_{2,2}+\Sigma_{Z_2}|}\geq\frac{|\Sigma_{S_2}+\Sigma_{Z_2}|}{|\Lambda_2+\Sigma_{Z_2}|}.\label{eq:Zsub2}
\end{align}
Substituting (\ref{eq:Zsub1}) and (\ref{eq:Zsub2}) into (\ref{eq:Pbound}) gives
\begin{align}
&P\geq\sup\limits_{\Sigma_Z\succ 0} N_1\Big(\frac{|\Sigma_S+\Sigma_Z|}{|\Lambda_1+\Sigma_{Z_1}||\Lambda_2+\Sigma_{Z_2}|}\Big)^{\frac{1}{\kappa}}\nonumber\\
&\hspace{0.7in}+(N_2-N_1)\Big(\frac{|\Sigma_{S_2}+\Sigma_{Z_2}|}{|\Lambda_2+\Sigma_{Z_2}|}\Big)^{\frac{1}{\kappa}}-N_2,\label{eq:tight}
\end{align}
where $\Sigma_Z$ is partitioned to the form in (\ref{eq:Zpar}). Setting $\kappa=1$ in (\ref{eq:tight}) recovers \cite[Corollary 1]{SCT15}. An equivalent form of the lower bound in (\ref{eq:tight}) was first obtained by Bross \textit{et al.} \cite{BLT10} via a different approach for the special case $\kappa=\ell_1=\ell_2=1$. It is worth mentioning that source-channel separation is known to be suboptimal in general for this problem \cite{TDS11,GT13}. Somewhat surprisingly, the lower bound in (\ref{eq:tight}),  derived with the aid of a source-channel separation theorem (i.e., Theorem \ref{thm:Gsep}), turns out to be tight when $\kappa=\ell_2=1$ \cite[Theorem 2]{SCT15} and is achievable by a class of hybrid digital-analog coding schemes\footnote{The hybrid scheme in \cite{TDS11} can be viewed as an extremal case of this class of schemes.} \cite[Section IV.B]{SCT15}. Therefore,  the application of source-channel separation theorems is not restricted to the relatively limited scenarios where the separation architecture is optimal; they can also be used to prove the optimality of non-separation based schemes and determine
the performance limits in certain scenarios where the separation architecture is suboptimal.




\section{Conclusion}\label{sec:conclusion}

We have established a source-channel separation theorem, which is further leveraged to derive a general necessary condition for the source broadcast problem.
It is intriguing to note that, in certain cases (see, e.g., Theorem \ref{thm:binary} and Theorem \ref{thm:Gchannel}), this necessary condition takes the form of comparison of two capacity regions. This is by no means a coincidence. In fact, it suggests a new direction that can be explored to establish stronger converse results  for the source broadcast problem \cite{KC15}.



\section*{Acknowledgment}

The authors would like to thank Prof. Chandra Nair for his valuable help.

\end{document}